\documentclass[12pt,letterpaper]{article}
\usepackage{graphics}
\usepackage{color}
\usepackage{booktabs}
\usepackage{cite}
\definecolor{grey}{rgb}{0.5,0.5,0.5}
\definecolor{dgreen}{rgb}{0.0,0.83,0.0}
\definecolor{dred}{rgb}{0.95,0.0,0.0}
\newcommand{\DRbar}{{\ensuremath{\overline{\mathrm{DR}}}}}
\newcommand{\MSbar}{{\ensuremath{\overline{\mathrm{MS}}}}}
\newcommand{\ttt}[1]{\texttt{#1}}
\newcommand{\tsc}[1]{\textsc{#1}}
\newcommand{\mrm}[1]{\ensuremath{\mathrm{#1}}}
\newcommand{\ti}[1]{\tilde{#1}}
\newcommand{\half}{\ensuremath{\textstyle \frac12}}
\newcommand{\st}{\ensuremath{^{\mathrm{st}}}}
\newcommand{\nd}{\ensuremath{^{\mathrm{nd}}}}
\newcommand{\rd}{\ensuremath{^{\mathrm{rd}}}}
\newcommand{\neut}{\ensuremath{\ti{\chi}^0}}

\newcommand{\charg}{\ensuremath{\ti{\chi}^+}}

\newcommand{\e}{\mathrm{e}}

\newcommand{\A}{\mathrm{A}}

\renewcommand{\H}{\mathrm{H}}

\newcommand{\Z}{{Z}}

\newcommand{\GeV}{\mathrm{GeV}}

\newcommand{\gbox}[1]{\colorbox{dgreen}{\small\vspace*{-3mm}\parbox{0.85cm}{\mbox{\textcolor{white}{\textsc{#1}}\hspace*{-2mm}}}\vspace*{-3mm}}}
\newcommand{\rbox}[1]{\colorbox{dred}{\small\vspace*{-3mm}\parbox{0.85cm}{\mbox{\textcolor{white}{\textsc{#1}}\hspace*{-2mm}}}\vspace*{-3mm}}}
\newcommand{\bbox}[1]{\colorbox{blue}{\vspace*{-3mm}\parbox{1.5cm}{\mbox{\textcolor{white}{\textsc{#1}}\hspace*{-2mm}}}\vspace*{-3mm}}}

\newlength{\captivewidth}
\newlength{\tablinsep}
\newlength{\halfpagewid}
\newlength{\halfpage}
\newlength{\abstwidth}
\newcommand{\snumentry}[2]{
\begin{minipage}[t]{1.2cm}\flushright\ttt{#1}\end{minipage}
\hspace{2mm}:
\begin{minipage}[t]{11cm}\noindent
#2\end{minipage}\\[1mm]
}
\newcommand{\numentry}[2]{
\begin{minipage}[t]{1.3cm}\flushright\ttt{#1}\end{minipage}
\hspace{5mm}: 
\begin{minipage}[t]{13cm}\noindent
#2\end{minipage}\\[2mm]
}
\newcommand{\entry}[1]{
\begin{minipage}[t]{1cm}\ \end{minipage} 
\begin{minipage}[t]{13.5cm}\noindent
#1\end{minipage}\\[2mm]
}

\long\def\symbolfootnote[#1]#2{\begingroup%
\def\thefootnote{\fnsymbol{footnote}}\footnote[#1]{#2}\endgroup}

\newcommand{\arrdes}[1]{\begin{center}\framebox{\parbox{\textwidth}{#1}}%
\end{center}}

\newcommand{\mgut}{\ensuremath{M_{\mathrm{input}}}}
\newcommand{\mmess}{\ensuremath{M_{\mathrm{mess}}}}

\newcommand{\rpv}{\ensuremath{\mathrm{RPV}}}

\renewcommand{\Re}[1]{\ensuremath{\mrm{Re}\left\{#1\right\}}}
\renewcommand{\Im}[1]{\ensuremath{\mrm{Im}\left\{#1\right\}}}
\def\beq{\begin{equation}}
\def\eeq{\end{equation}}
\def\bea{\begin{eqnarray}}
\def\eea{\end{eqnarray}}

\begin{document}
\vspace*{-12mm}\begin{minipage}{0.98\textwidth}\tiny
\flushright
FERMILAB-PUB-07-036-T,\\
CERN-PH-TH/2007-148,DAMTP-2007-76, \\
Edinburgh 2007/31, KEK-TH-1170,LAPTH-1204/07, \\ 
LPT-ORSAY-07-81, SHEP-07-13, SLAC-PUB-12765.
\end{minipage}\\[8mm]\hspace*{-0.025\textwidth}
\begin{center}
\Large{\bf
SUSY Les Houches Accord 2
}\\[9mm]
\normalsize B.C.~Allanach$^1$, C.~Bal\'azs$^2$, G.~B\'elanger$^3$,
 M.~Bernhardt$^4$, F.~Boudjema$^3$, D. Choudhury$^5$, K.~Desch$^4$,
 U.~Ellwanger$^6$, P.~Gambino$^7$, R.~Godbole$^8$, T.~Goto$^9$, 
J.~Guasch$^{10}$,
 M.~Guchait$^{11}$, T.~Hahn$^{12}$, S.~Heinemeyer$^{13}$,
 C.~Hugonie$^{14}$, T.~Hurth$^{15,16}$, S.~Kraml$^{15,17}$,
 S.~Kreiss$^{18}$, J.~Lykken$^{19}$, 
 F.~Moortgat$^{20}$, S.~Moretti$^{6,21}$, S.~Pe\~naranda$^{22}$,
 T.~Plehn$^{18}$, W.~Porod$^{23}$, A.~Pukhov$^{24}$,
 P.~Richardson$^{15,25}$, M.~Schumacher$^{26}$, L.~Silvestrini$^{27}$,
 P.~Skands$^{15,19}$\symbolfootnote[1]{Corresponding author: Peter
   Skands (peter.skands@cern.ch), CERN PH-TH, Geneva 23, CH-1211,
   Switzerland; tlph: +41-2276-72447; see
 \texttt{http://home.fnal.gov/$\sim$skands/slha/} for updates and
 examples.}, 
 P.~Slavich$^{3,15}$, M.~Spira$^{28}$, G.~Weiglein$^{25}$,
 P.~Wienemann$^{4}$ \\[5mm]
\begin{minipage}{14.5cm}
\begin{abstract}
The Supersymmetry Les Houches Accord (SLHA)
provides a universal set of conventions
for conveying spectral and decay information for supersymmetry 
analysis problems in high energy physics. 
Here, we propose extensions of the conventions of the first SLHA 
to include various generalisations: the minimal supersymmetric
standard model with violation of CP, R-parity, and
flavour, as well as the simplest next-to-minimal model. 
\end{abstract}
\end{minipage}
\\[10mm]
{\it\footnotesize
\begin{tabular}{rp{15cm}}
1 & DAMTP, CMS, University of Cambridge, Wilberforce Road, Cambridge, CB3 0WA, United Kingdom \\
2 & School of Physics, Monash University, Melbourne VIC 3800, Australia \\
3 & LAPTH, CNRS, UMR 5108, Chemin de Bellevue BP110, F-74941, Annecy-le-Vieux, France \\
4 & University of Bonn, Physics Institute, Nussallee 12, D-53115 Bonn, Germany\\
5 & Harish Chandra Research Institute, Chhatnag Road, Jhunsi, Allahabad
211019, India\\
6 & Laboratoire de Physique Th\'{e}orique, Universit\'{e} de Paris Sud XI,
B\^{a}t.~210, F-91405, Orsay, France\\
7 & Universit\`a di Torino and INFN, sez.\ di Torino, 10125 Torino,
Italy\\
8 & Indian Institute of Science,
Center for Theoretical Studies,
Bangalore 560012, Karnataka,
India\\
9 & Theory Group, IPNS, KEK, Tsukuba, 305-0801, Japan\\
10 & Universitat de Barcelona,
Dept.\ de Fisica Fonamental,
Diagonal 647,
E-08028 Barcelona, Spain\\
11 & Tata Institute for Fundamental Research,
Homi Bhabha Road,
Mumbai,
India\\
12 & MPI, Werner Heisenberg Inst., F\"ohringer Ring 6, 
D-80805 Munich, Germany\\
13 & IFCA (CSIC--UC), Avda. de los Castros s/n, 39005 Santander,
Spain\\
14 &LPTA, Universit\'{e} Montpellier II,
F-34095 Montpellier, Cedex 5,
France\\
15 & 
CERN, Department of Physics, Theory Unit, CH-1211, Geneva 23, 
Switzerland\\
16 & SLAC, Stanford University, Stanford, CA 94309, USA\\
17 & Laboratoire de Physique Subatomique et de Cosmologie (LPSC), UJF
Grenoble 1, CNRS/IN2P3, 53 Avenue des Martyrs, F-38026 Grenoble, France\\
18 & The University of Edinburgh,
School of Physics,
James Clerk Maxwell Building, King's Buildings,
Mayfield Road,
Edinburgh EH9 3JZ, United Kingdom\\
19 & Theoretical Physics, Fermilab,
P.O. Box 500,
Batavia, IL 60510-0500,
USA\\
\end{tabular}
\begin{tabular}{rp{13.3cm}}
20 & ETH Zurich, CH-8093 Zurich, Switzerland \\
21 & University of Southampton,
School of Physics and Astronomy,
Highfield,
Southampton S017 1BJ, United Kingdom\\
22 & Departamento de F{\'{\i}}sica Te{\'o}rica, Universidad
de Zaragoza, E-50009, Zaragoza, Spain \\
23 & Institut f\"ur Theoretische Physik und Astrophysik,
  Universit\"at W\"urzburg, 97074 W\"urzburg, Germany\\
24 & Skobeltsyn Inst.\ of Nuclear Physics, Moscow State Univ., 
Moscow 119991, Russia \\
25 & Durham University
Inst. for Particle Physics Phenomenology
Ogden Centre for Fundamental Physics
South Road
Durham DH1 3LE, United Kingdom\\
26 & Univ. G\"ottingen,
Inst.\ f\"ur Theoretische Physik,
Friedreich-Hund-Platz 1,
D-37077 G\"ottingen, Germany\\
27 & Univ.\ degli Studi di Roma, La Sapienza,
Dipt.\ di Fisica G.\ Marconi,
Piazzale Aldo Moro 2,
I-00185 Rome, Italy\\
28 & Paul Scherrer Institute,
5232 Villigen PSI,
Switzerland \\
\end{tabular}}\\
\end{center}
\hrule
\tableofcontents
\newpage

\section{Introduction \label{sec:intro}}
Supersymmetric (SUSY) extensions of the Standard Model rank
among the most promising and well-explored scenarios for New Physics
at the TeV scale. Given the long history of supersymmetry 
and the number of people working in the field, several
different conventions for
defining supersymmetric theories have been proposed over the years,
many of which have come into widespread use. 
At present, therefore, no unique set of conventions prevails.
In principle, this is not a problem. As long as
everything is clearly and consistently defined, a translation can always
be made between two sets of conventions.

However, the proliferation of conventions does have 
some disadvantages. Results obtained by different authors or computer codes
are not always directly comparable. Hence, if author/code A wishes to
use the results of author/code B in a calculation, a consistency check
of all the relevant conventions and any necessary translations must first be
made -- a tedious and error-prone task.

To deal with this problem, and to create a more transparent situation
for non-experts, the original SUSY Les Houches Accord (SLHA1) was proposed
\cite{Skands:2003cj}. This accord uniquely defines a 
set of conventions for supersymmetric models together with a common interface
between codes. The most essential fact is not what the conventions are
in detail (they largely resemble those of
\cite{Chung:2003fi}), but that 
they are consistent and unambiguous, hence reducing the
problem of translating between conventions to a linear, rather than a
factorial, dependence on the number of codes involved. 
At present, these codes can be categorised roughly as follows (see
\cite{Allanach:2008zn,bsmrepository} for a review and on-line repository): 
\begin{itemize}
\item Spectrum
calculators~\cite{Baer:1999sp,Allanach:2001kg,Porod:2003um,Djouadi:2002ze},
which calculate 
the supersymmetric mass and coupling spectrum, 
assuming some (given or derived) 
SUSY-breaking terms and a matching to known data on the Standard
Model parameters.
\item
Observables
calculators~\cite{Beenakker:1996ed,Djouadi:1998yw,Heinemeyer:1998yj,Muhlleitner:2003vg,Lee:2003nta,Belanger:2004yn,Gondolo:2004sc,Ellwanger:2004xm,fchdecay,Degrassi:2007kj,Mahmoudi:2007vz}; 
packages which calculate one or more of the following: 
collider production cross sections (cross section calculators), 
decay partial widths (decay packages), relic dark matter density (dark
matter packages), and  indirect/precision 
observables,  
such as rare decay branching ratios or Higgs/electroweak observables
(constraint packages).   
\item
Monte-Carlo event
generators~\cite{Corcella:2000bw,Moretti:2002eu,Sjostrand:2006za,Mrenna:1996hu,Hagiwara:2005wg,Katsanevas:1998fb,Reuter:2005us,Cho:2006sx,Kilian:2007gr},    
which calculate cross sections through explicit statistical simulation
of high-energy particle collisions. By including 
resonance decays, parton showering,
hadronisation, and underlying-event effects, fully
exclusive final states can be studied,
and, for instance, detector simulations interfaced.
\item
SUSY and CKM fitting
programs~\cite{Bechtle:2004pc,Lafaye:2004cn,Ciuchini:2000de,Charles:2004jd} 
which fit model parameters to collider-type data.
\end{itemize}\vskip3mm

At the time of writing, the SLHA1 has already, to a large extent,
obliterated the need for separately coded (and
maintained and debugged)  
interfaces between many of these codes. Moreover, it has provided users with
input and output in a common format, which is more readily comparable
and transferable. Finally, the SLHA convention choices are also being 
adapted for other tasks, such as the SPA project
\cite{Aguilar-Saavedra:2005pw}. We believe, therefore, that the SLHA
project has been useful, solving a problem that, for experts, is trivial but
frequently occurring and tedious to deal with, and which, for
non-experts, is an unnecessary headache. 

However, SLHA1 was designed exclusively with the MSSM with real
parameters and R-parity conservation in mind. Some recent public
codes~\cite{Dreiner:1999qz,Skands:2001it,Allanach:2001kg,Sjostrand:2002ip,Porod:2003um,Ellwanger:2005dv,Frank:2006yh,fchdecay,Degrassi:2007kj}
are either implementing extensions to this base model or are
anticipating such extensions. It therefore seems prudent at this time
to consider how to extend SLHA1 to deal with more general
supersymmetric theories. In particular, we will consider R-parity
violation (\rpv), flavour violation (FLV), and CP-violating (CPV)
phases in the minimal supersymmetric standard model (MSSM).  We will
also consider next-to-minimal models (i.e., models in which the MSSM
field content is augmented by a gauge-singlet chiral superfield) which
we shall collectively label by the acronym NMSSM.

Rather than giving exhaustive historical references for all concepts
used in this article, we provide a list of useful and pedagogical
reviews to whose contents and references in turn we refer. For the 
various topics treated in the article, these reviews are:
\begin{itemize}
\item SUSY \cite{Martin:1997ns},
\item FLV \cite{Misiak:1997ei},
\item Neutrinos \cite{GonzalezGarcia:2007ib},
\item \rpv \cite{Barbier:2004ez,Hirsch:2000ef},
\item CPV \cite{CPV-Accomando:2006ga}, 
\item NMSSM \cite{NMSSM-Accomando:2006ga}, 
\item SUSY Tools \cite{Allanach:2008zn,bsmrepository}.
\end{itemize}

There is clearly some tension between the desirable goals of
generality of the models, ease of implementation in programs, 
and practicality for users. 
A completely general accord would be useless in practice if it was 
so complicated that no one would implement it. 
We have agreed on the following for SLHA2:
for the MSSM, we will here restrict our attention to \emph{either} CPV or
\rpv, but not both. For RPV and flavour violation, 
we shall work in the Super-CKM/PMNS basis, 
as defined in sections \ref{sec:flv} and \ref{sec:rpv}. 
For the NMSSM, we define one catch-all model and extend the SLHA1 mixing only 
to include the new states, with CP, R-parity, and flavour still
assumed conserved.  

To make the interface independent of programming languages, compilers,
platforms etc, the SLHA1 is based on the transfer of three different
ASCII files (or potentially a character string containing identical
ASCII information): one for model input, one for spectrum calculator
output, and one for decay calculator output.  We believe that the
advantage of implementation independence outweighs the disadvantage of
codes using SLHA1 having to parse input. Indeed, there are tools to
assist with this task~\cite{Hahn:2004bc,Hahn:2006nq,PlehnAndKreiss}.

Care was taken in SLHA1 to provide a framework for the MSSM that could
easily be extended to the cases listed above. The conventions and
switches described here are designed to be a {\em superset} of those
of the original SLHA1 and so, unless explicitly mentioned in the text,
we will assume the conventions of the original
accord~\cite{Skands:2003cj} implicitly.  For instance, all dimensionful
parameters quoted in the present paper are assumed to be in the
appropriate power of GeV, all angles are in radians, and the output
formats for SLHA2 data \ttt{BLOCK}s follow those of SLHA1.  In a few
cases it will be necessary to replace the original conventions. This
is clearly remarked upon in all places where it occurs, and the SLHA2
conventions then supersede the SLHA1 ones.

\section{Extensions of SLHA1 \label{sec:slha1}}
Since its first publication, a few useful extensions to the
SLHA1 have been identified. These are collected here for reference and
are independent of the more general SUSY models discussed in
subsequent sections. (Also note the recent proposal for a joint
SLHA+LHEF format for BSM event generation \cite{Alwall:2006yp,Alwall:2007mw}.)

Firstly, we introduce additional optional
entries in the SLHA1 block \ttt{EXTPAR} to allow for using 
either the $A^0$ or $H^+$ pole masses as input instead of the parameter
$m_A^2(\mgut)$ defined in \cite{Skands:2003cj}. 
 
Secondly, to allow for
different parameters to be defined at different scales (e.g., $\mu$
defined at $M_{\mrm{EWSB}}$, the remaining parameters defined at
$\mgut$) we introduce a new optional block
\ttt{QEXTPAR} which, if present, overrides the default \ttt{MINPAR}
and \ttt{EXTPAR} scale choices for specific parameters, as defined
below. 

While there is no obligation on codes to implement these
extensions,  we perceive it as useful that the accord
allows for them, enabling a wider range of input parameter sets to be
considered. The entries defined in \ttt{EXTPAR} and \ttt{QEXTPAR}
in the SLHA2 are thus (repeating unchanged \ttt{EXTPAR} entries for
completeness): 
\subsection*{\ttt{BLOCK EXTPAR}}
Optional input parameters for non-minimal/non-universal
models. This block may be entirely absent from the input file, in which case
a minimal type of the selected SUSY breaking model will be used. When
block \ttt{EXTPAR} is present, the starting
point is still a minimal model with parameters as given in
\ttt{MINPAR} \cite{Skands:2003cj} but 
with each value present in \ttt{EXTPAR} replacing the 
minimal model value of that parameter, as applicable. If
\ttt{MINPAR} is not present, then \emph{all} model parameters must be
specified explicitly using \ttt{EXTPAR}. All scale-dependent parameters are
understood to be given in the \DRbar\ scheme.\\[3mm]
\clearpage
\arrdes{Input scale}
\numentry{0}{$\mgut$. Input scale for \ttt{EXTPAR} entries in 
SUGRA, AMSB, and general MSSM models. 
If absent, the GUT scale derived from gauge unification will be used as input
scale. Note that this parameter has no effect in GMSB scenarios
  where the input scale by definition is identical to the messenger scale,
  $\mmess$. A special case is when 
  $Q=M_{\mrm{EWSB}}\equiv\sqrt{m_{\ti{t}_1}m_{\ti{t}_2}}$ is desired as input
  scale, since this scale is not known beforehand. This choice can be
  invoked by giving the special value $\mgut=-1$. To define an
  alternative input scale for one or more specific parameters, see
  \ttt{QEXTPAR} below. 
}
\arrdes{Gaugino Masses}
\numentry{1}{$M_1(\mgut)$. $U(1)_Y$ gaugino (Bino) mass.} 
\numentry{2}{$M_2(\mgut)$. $SU(2)_L$ gaugino (Wino) mass.}
\numentry{3}{$M_3(\mgut)$. $SU(3)_C$ gaugino (gluino) mass.} 
\arrdes{Trilinear Couplings}
\numentry{11}{$A_t(\mgut)$. Top trilinear coupling. }
\numentry{12}{$A_b(\mgut)$. Bottom trilinear coupling.}
\numentry{13}{$A_\tau(\mgut)$. Tau trilinear coupling.}
\arrdes{Higgs Parameters\\[1mm]
{--- Only one of the parameter sets ($m_{H_1}^2$,
  $m_{H_2}^2$), ($\mu,m_A^2$), ($\mu,m_{\A^0}$), or ($\mu,m_{\H^+}$) 
should be given, they merely represent different ways of specifying
the same parameters.}} 
\numentry{21}{$m_{H_1}^2(\mgut)$. Down type Higgs mass squared.} 
\numentry{22}{$m_{H_2}^2(\mgut)$. Up type Higgs mass squared. }
\numentry{23}{$\mu(\mgut)$. $\mu$ parameter.}
\numentry{24}{$m_A^2(\mgut)$. Tree--level pseudoscalar Higgs mass
  parameter squared, as defined in \cite{Skands:2003cj}.}
\numentry{25}{$\tan\beta(\mgut)$. If present, this value of $\tan\beta$
  overrides the one in \ttt{MINPAR}, and the input scale is taken as $\mgut$
  rather than $m_\Z$.}
\numentry{26}{$m_{\A^0}$. 
  Pseudoscalar Higgs pole mass. May be given
  instead of $m_A^2(\mgut)$.}
\numentry{27}{$m_{\H^+}$. 
  Charged Higgs pole mass. May be given
  instead of $m_A^2(\mgut)$.}
\arrdes{Sfermion Masses}
\numentry{31}{$m_{\ti{e}_L}(\mgut)$. Left 1\st gen.\ scalar lepton mass. }
\numentry{32}{$m_{\ti{\mu}_L}(\mgut)$. Left 2\nd gen.\ scalar lepton mass.}
\numentry{33}{$m_{\ti{\tau}_L}(\mgut)$. Left 3\rd gen.\ scalar lepton mass. }
\numentry{34}{$m_{\ti{e}_R}(\mgut)$. Right scalar electron mass.}
\numentry{35}{$m_{\ti{\mu}_R}(\mgut)$. Right scalar muon mass.}
\numentry{36}{$m_{\ti{\tau}_R}(\mgut)$. Right scalar tau mass.}
\numentry{41}{$m_{\ti{q}_{1L}}(\mgut)$. Left 1\st gen.\ scalar quark mass. }
\numentry{42}{$m_{\ti{q}_{2L}}(\mgut)$. Left 2\nd gen.\ scalar quark mass. }
\numentry{43}{$m_{\ti{q}_{3L}}(\mgut)$. Left 3\rd gen.\ scalar quark mass. }
\numentry{44}{$m_{\ti{u}_{R}}(\mgut)$. Right scalar up mass. }
\numentry{45}{$m_{\ti{c}_{R}}(\mgut)$. Right scalar charm mass. }
\numentry{46}{$m_{\ti{t}_{R}}(\mgut)$. Right scalar top mass. }
\numentry{47}{$m_{\ti{d}_{R}}(\mgut)$. Right scalar down mass. }
\numentry{48}{$m_{\ti{s}_{R}}(\mgut)$. Right scalar strange mass. }
\numentry{49}{$m_{\ti{b}_{R}}(\mgut)$. Right scalar bottom mass. }
\arrdes{Other Extensions}
\numentry{51}{$N_1$ (GMSB only). 
$U(1)_Y$ messenger index (defined as in ref.~\cite{Martin:1996zb}).}
\numentry{52}{$N_2$ (GMSB only). 
$SU(2)_L$ messenger index (defined as in ref.~\cite{Martin:1996zb}).}
\numentry{53}{$N_3$ (GMSB only). 
$SU(3)_C$ messenger index (defined as in ref.~\cite{Martin:1996zb}).}

\subsection*{\ttt{BLOCK QEXTPAR}}
Optional alternative input scales for specific parameters. This
block should normally be absent, in which case 
the default input scale or $\mgut$ (see \ttt{EXTPAR 0}) 
will be used for all parameters. We stress that most codes cannot be
expected to allow for multiple arbitrary scale choices, 
so the relevant manual and output should be carefully checked
to make sure the desired behaviour is obtained. Currently defined
entries are:\\[3mm]
\numentry{1}{$Q_{M_1}$. Input scale for $M_1$.} 
\numentry{2}{$Q_{M_2}$. Input scale for $M_2$.}
\numentry{3}{$Q_{M_3}$. Input scale for $M_3$.} 
\numentry{11}{$Q_{A_u}$. Input scale for up-type squark trilinear couplings.}
\numentry{12}{$Q_{A_d}$. Input scale for down-type squark trilinear couplings.}
\numentry{13}{$Q_{A_\ell}$. Input scale for charged slepton trilinear
  couplings.} 
\numentry{21}{$Q_{m_{H_1}^2}$. Input scale for $m_{H_1}^2$.} 
\numentry{22}{$Q_{m_{H_2}^2}$. Input scale for $m_{H_2}^2$.} 
\numentry{23}{$Q_{\mu}$. Input scale for $\mu$.}
\numentry{24}{$Q_{m_A^2}$. Input scale for $m_A^2$, as defined in
  \cite{Skands:2003cj}.} 
\numentry{25}{$Q_{\tan\beta}$. Input scale for $\tan\beta$.}
\numentry{31}{$Q_{m_{\ti{\ell}_L}}$. Input scale for all left-handed
    slepton mass terms.}
\numentry{34}{$Q_{m_{\ti{\ell}_R}}$. Input scale for all right-handed 
    slepton mass terms.} 
\numentry{41}{$Q_{m_{\ti{q}_{L}}}$. Input scale for all left-handed
    squark mass terms.}  
\numentry{44}{$Q_{m_{\ti{u}_{R}}}$.  Input scale for all right-handed
    up-type squark mass terms.} 
\numentry{47}{$Q_{m_{\ti{d}_{R}}}$.  Input scale for all right-handed
    down-type squark mass terms.}

\section{Model Selection \label{sec:modelselection}}
To define the general properties of the model, we propose to introduce
global switches in the SLHA1 model definition block \ttt{MODSEL}, as
follows.  Note that the switches defined here are in addition to the
ones in \cite{Skands:2003cj}.

\subsection*{\ttt{BLOCK MODSEL}\label{sec:modsel}}
Switches and options for model selection. The entries in this block
should consist of an index, identifying the particular switch in the
listing below, followed by another integer or real number, specifying
the option or value chosen: \\[2mm]
\numentry{3}{(Default=\ttt{0}) Choice of particle content.
Switches defined are:\\
\snumentry{0}{MSSM. This corresponds to SLHA1.}
\snumentry{1}{NMSSM. The blocks defined in section \ref{sec:nmssm}
  should be present.}
}
\numentry{4}{(Default=\ttt{0}) R-parity violation. Switches defined are:\\
\snumentry{0}{R-parity conserved. This corresponds to the SLHA1.}
\snumentry{1}{R-parity violated. The blocks defined in
  section~\ref{sec:rpv} should be present.} 
} 
\numentry{5}{(Default=\ttt{0}) CP violation. Switches defined are:\\
\snumentry{0}{CP is conserved. No information even on the CKM phase
is used. This corresponds to the SLHA1.}
\snumentry{1}{CP is violated, but only by the standard CKM
phase. All other phases are assumed zero.}
\snumentry{2}{CP is violated. Completely general CP phases
allowed. Imaginary parts corresponding to the entries in the SLHA1 block
\ttt{EXTPAR} can be given in \ttt{IMEXTPAR} (together with the CKM
phase). In the case of additional SUSY flavour violation,
imaginary parts of the blocks defined in section~\ref{sec:flv}
should be given, again with the prefix \ttt{IM}, 
which supersede the corresponding entries in \ttt{IMEXTPAR}.}
}
\numentry{6}{(Default=\ttt{0}) Flavour violation. Switches defined are:\\
\snumentry{0}{No (SUSY) flavour violation. This corresponds to the
SLHA1. 
}
\snumentry{1}{Quark flavour is violated. The blocks defined in section
\ref{sec:flv} should be present.}
\snumentry{2}{Lepton flavour is violated. The blocks defined in section
\ref{sec:flv} should be present.}
\snumentry{3}{Lepton and quark flavour is violated. The blocks defined in
  section \ref{sec:flv} should be present.}
}

\section{General MSSM}
For convenience, we here repeat the definitions of the field content,
superpotential and soft SUSY-breaking potential of the MSSM in the
notation of \cite{Skands:2003cj}.

Specifically, 
the chiral superfields of the MSSM have the following $SU(3)_C\otimes
SU(2)_L\otimes U(1)_Y$ quantum numbers
\begin{eqnarray}
L:&(1,2,-\half),\quad {\bar E}:&(1,1,1),\qquad\, \textstyle
Q:\,(3,2,\frac16),\quad
{\bar U}:\,(\bar{3},1,-\frac{2}{3}),\nonumber\\ {\bar D}:&(\bar{3},1,\frac13),\quad
H_1:&(1,2,-\half),\quad  H_2:\,(1,2,\half)~,
\label{fields}
\end{eqnarray}
the superpotential 
(omitting RPV terms, see section \ref{sec:rpv}) is written as
\begin{eqnarray}
W_{MSSM}&=& \epsilon_{ab} \left[ 
  (Y_E)_{ij} H_1^a    L_i^b    {\bar E}_j 
+ (Y_D)_{ij} H_1^a    Q_i^{b}  {\bar D}_{j} 
+ (Y_U)_{ij} H_2^b    Q_i^{a}  {\bar U}_{j}  
- \mu H_1^a H_2^b \right]~,
\label{eq:wmssm}
\end{eqnarray}
and the trilinear and bilinear soft SUSY-breaking potentials 
$V_3$ and $V_2$ are
\begin{eqnarray}
V_3 & = & \epsilon_{ab} \sum_{ij}
\left[
(T_E)_{ij} H_1^a \tilde{L}_{i_L}^{b} \tilde{e}_{j_R}^* +
(T_D)_{ij} H_1^a               \tilde{Q}_{i_L}^{b}  \tilde{d}_{j_R}^* +
(T_U)_{ij}  H_2^b \tilde{Q}_{i_L}^{a} \tilde{u}_{j_R}^*
\right]
+ \mrm{h.c.}~, \label{eq:slha1v3soft}\\[2mm]
V_2 &=& m_{H_1}^2 {{H^*_1}_a} {H_1^a} + m_{H_2}^2 {{H^*_2}_a} {H_2^a} +
{\tilde{Q}^*}_{i_La} (m_{\tilde Q}^2)_{ij} \tilde{Q}_{j_L}^{a} +
{\tilde{L}^*}_{i_La} (m_{\tilde L}^2)_{ij} \tilde{L}_{j_L}^{a}  
+ \nonumber \\ &&
\tilde{u}_{i_R} (m_{\tilde u}^2)_{ij} {\tilde{u}^*}_{j_R} +
\tilde{d}_{i_R} (m_{\tilde d}^2)_{ij} {\tilde{d}^*}_{j_R} +
\tilde{e}_{i_R} (m_{\tilde e}^2)_{ij} {\tilde{e}^*}_{j_R} -
(m_3^2 \epsilon_{ab} H_1^a H_2^b + \mrm{h.c.})~, \label{eq:slha1v2soft}
\end{eqnarray}
where a tilde over the symbol for a quark or lepton superfield denotes 
its scalar component  (note however that we define, e.g.,
$\tilde{u}_R^*$ as the scalar component of $\bar{U}$).
Throughout this section, we denote $SU(2)_L$ fundamental representation
indices by $a,b=1,2$ and generation indices by $i,j=1,2,3$. Colour indices
are everywhere suppressed, since only trivial contractions are involved.
$\epsilon_{ab}$ is the totally antisymmetric tensor, with
$\epsilon_{12}=\epsilon^{12}=1$. 

\subsection{Flavour Violation \label{sec:flv}}

\subsubsection{The quark sector and the super-CKM basis}
\label{flavchange}
Within the MSSM there are in
general new sources of flavour violation arising from a possible
misalignment of quarks and squarks in flavour space. 
The severe experimental
 constraints on flavour violation have no direct explanation in the
 structure of the unconstrained MSSM which leads to the well-known
 supersymmetric flavour problem.

The Super-CKM basis of the squarks 
is very useful in this context because in that basis  
only physically measurable parameters are present. 
In the Super-CKM basis the quark mass matrix is diagonal and the
squarks are rotated in parallel to their superpartners. 
Actually, once  the electroweak symmetry is broken, a rotation in 
flavour space 
\begin{equation}
  D^{\,o}     \  = \ V_d \,D\,,           \hspace*{0.8truecm}
  U^{\,o}     \  = \ V_u \,U\,,           \hspace*{0.8truecm}
  \bar{D}^{o}     = \ U_d^\ast \,\bar{D}\,, \hspace*{0.8truecm}
  \bar{U}^{o}       = \ U_u^\ast \,\bar{U}\,, 
\label{superotation}
\end{equation}
of all matter superfields in the (s)quark superpotential 
\begin{equation}
 W_Q  \ = \ \epsilon_{ab}  
  \left[ \left(Y_D\right)_{ij}  H_1^a Q_i^{b\,o} \bar{D}_j^{\,o}
  +    \left(Y_U\right)_{ij} H_2^b  Q_i^{a\,o}  \bar{U}_j^{\,o}
 \, \right]
\label{superpot}
\end{equation}
brings fermions from the interaction eigenstate basis 
$\{d_L^o,u_L^o,d_R^o,u_R^o\}$ to their mass eigenstate basis 
$\{d_L,u_L,d_R,u_R\}$:  
\begin{equation}
d_L^o = V_d d_L\,, \hspace*{0.8truecm}
u_L^o = V_u u_L\,, \hspace*{0.8truecm}
d_R^o = U_d d_R\,, \hspace*{0.8truecm}
u_R^o = U_u u_R\,, 
\label{fermrotation}
\end{equation}
and the scalar superpartners to the basis 
$\{ \tilde{d}_L, \tilde{u}_L, 
    \tilde{d}_R, \tilde{u}_R \}$.
Through this rotation, the Yukawa matrices $Y_D$ and $Y_U$ are 
reduced to their diagonal form $\hat{Y}_D$ and $\hat{Y}_U$: 
\begin{equation}
(\hat{Y}_D)_{ii} = (U_d^\dagger Y_D^T V_d)_{ii}= \sqrt{2}\frac{m_{d\,i}}{v_1}\,,
\hspace*{1.0truecm} 
(\hat{Y}_U)_{ii} = (U_u^\dagger Y_U^T V_u)_{ii}=
\sqrt{2}\frac{m_{u\,i}}{v_2}\,. 
\label{diagyukawa}
\end{equation}
Tree-level mixing terms among quarks of different generations are due 
to the misalignment of $V_d$ and $V_u$, expressed via the CKM matrix 
\begin{equation}
V_{\mrm{CKM}}=V_u^\dagger V_d~,
\end{equation}
which is proportional to the tree-level
$\bar{u}_{Li}d_{Lj}W^+ $, 
$\bar{u}_{Li}d_{Rj}H^+ $, and 
$\bar{u}_{Ri}d_{Lj}H^+$ couplings ($i,j=1,2,3$). 
This is also true for the supersymmetric counterparts of these vertices, 
in the limit of unbroken supersymmetry.

In the super-CKM  basis the $6\times 6$ mass matrices for the up-type and
down-type squarks are defined as
\begin{equation}
{\cal L}^{\rm mass}_{\tilde q} ~=~ 
- \Phi_u^{\dagger}\,
{\cal M}_{\tilde u}^2\, 
\Phi_u
- \Phi_d^{\dagger}\,
{\cal M}_{\tilde d}^2\, 
\Phi_d~,
\end{equation}
where $\Phi_u = (\tilde u_L,\tilde c_L, \tilde t_L,
\tilde u_R,\tilde c_R, \tilde t_R)^T$ and 
$\Phi_d = (\tilde d_L,\tilde s_L, \tilde b_L,
\tilde d_R,\tilde s_R, \tilde b_R)^T$.
We diagonalise the squark mass matrices via $6\times6$ unitary
matrices $R_{u,d}$, such that $R_{u,d} \,{\cal M}_{{\tilde
    u},{\tilde d}}^2\,R_{u,d}^\dagger$ are diagonal matrices with increasing
mass squared values. 
The flavour-mixed mass matrices read:
\begin{eqnarray}
{\cal M}_{\tilde u}^2 &=&  \left( \begin{array}{cc}
  V_{\mrm{CKM}} \,{\hat{m}_{\tilde Q}}^2\, V_{\mrm{CKM}}^\dagger + m^2_{u}
  + D_{u\,LL} 
  &   
  \frac{v_2}{\sqrt{2}} {\hat T}_U^\dagger  - \mu m_u \cot\beta      \\[1.01ex]
  \frac{v_2}{\sqrt{2}} {\hat T}_U  - \mu^* m_u \cot\beta  &
       {\hat m}^2_{\tilde u} + m^2_{u} + D_{u\,RR} \\                 
 \end{array} \right) \,\, ,
\label{massmatrixu}  \\
\nonumber\\
{\cal M}_{\tilde d}^2 &=&  \left( \begin{array}{cc}
   {\hat m_{\tilde Q}}^2 + m^2_{d} + D_{d\,LL}           
& 
   \frac{v_1}{\sqrt{2}} {\hat T}_D^\dagger  - \mu m_d \tan\beta      \\[1.01ex]
   \frac{v_1}{\sqrt{2}} {\hat T}_D  - \mu^* m_d \tan\beta  &
   {\hat m}^2_{\tilde d} + m^2_{d} + D_{d\,RR}
             \\                 
 \end{array} \right) \,\, .
\label{massmatrixd}  
\end{eqnarray}
In the equations above we introduced the $3\times 3$ matrices
\begin{equation}
{\hat m_{\tilde Q}}^2 \equiv V^\dagger_d \,m^2_{\tilde Q}\, V_d\,,~~~
{\hat m_{\tilde u}}^2 \equiv U^\dagger_u \,{m^2_{\tilde u}}^T\, U_u\,,~~~
{\hat m_{\tilde d}}^2 \equiv U^\dagger_d \,{m^2_{\tilde d}}^T\, U_d\,,
\label{eq:mhat}
\end{equation}
\begin{equation}
{\hat T_{U}} \equiv U^\dagger_u \,T_{U}^T\, V_u\,,~~~
{\hat T_{D}} \equiv U^\dagger_d \,T_{D}^T\, V_d\,,~~~
\label{eq:that}
\end{equation}
where the un-hatted mass matrices $m^2_{Q,u,d}$ 
and trilinear interaction matrices $T_{U,D}$ 
are given in the interaction basis.

The
matrices $m_{u,d}$ are the diagonal up-type and down-type quark masses
and $D_{f\,LL,RR}$ are the D-terms given by:
 \begin{equation}
 D_{f\,LL,RR} =  \cos 2\beta \, m_Z^2 
   \left(T_f^3 - Q_f \sin^2\theta_W \right) 
{{\mathchoice {\rm 1\mskip-4mu l} {\rm 1\mskip-4mu l}
{\rm 1\mskip-4.5mu l} {\rm 1\mskip-5mu l}}}_3\,,
\label{dterm}
\end{equation}
which are also flavour diagonal. Here, $Q_f$ is the electric
charge of the left-handed chiral supermultiplet to which the squark
belongs, i.e., it is $2/3$ for $U$ and $-2/3$ for $U^c$. 
Note that the up-type and down-type
squark mass matrices in eqs.~(\ref{massmatrixu}) and
(\ref{massmatrixd}) cannot be simultaneously flavour-diagonal unless
${\hat m_{\tilde Q}}^2$ is flavour-universal (i.e., proportional to the
identity in flavour space).

\subsubsection{The lepton sector and the super-PMNS basis \label{sec:lep}}
For the lepton sector, we adopt a super-PMNS basis, as defined in this
section.

Neutrino oscillation data have provided a strong indication
that neutrinos have masses and that there are flavour-changing charged
currents in the leptonic sector. One popular model to produce such effects
is the see-saw mechanism, where right-handed neutrinos have both 
Majorana masses as well as Yukawa couplings with the left-handed
leptons.
When the heavy neutrinos
are integrated out of the effective field theory, one is left with three light
approximately left-handed neutrinos which are identified with the ones
observed experimentally. There are other models of neutrino masses, for
example involving SU(2) Higgs triplets, that, once the triplets have been
integrated out, also lead to effective Majorana masses for the neutrinos.
Here, we cover all cases that lead to a low energy effective field theory
with Majorana neutrino masses and one sneutrino per family. 
In terms of this low energy effective theory, the lepton mixing phenomenon is
analogous to the quark mixing case and so we adapt the conventions defined
above to the leptonic case.

After electroweak symmetry breaking, the
neutrino sector of the MSSM contains the Lagrangian pieces (in
2--component notation)
\begin{equation}
{\mathcal L} = -\frac12 {\nu}^{oT} (m_\nu) \nu^o
+ {\rm h.c.},
\end{equation}
where $m_\nu$ is a $3 \times 3$ symmetric matrix. The 
interaction eigenstate basis neutrino fields $\nu^o$ are related to the mass
eigenstate ones $\nu$ by
\begin{equation}
\nu^o = V_\nu \nu,
\end{equation}
reducing the mass matrix $m_\nu$ 
to its diagonal form $\hat{m}_\nu$ 
\begin{equation}
  (\hat{m}_\nu)_{ii} = (V_\nu^T m_\nu V_\nu)_{ii} =
  m_{\nu_i}. \label{eq:diagmL} 
\end{equation}
The charged
lepton fields have a $3 \times 3$ Yukawa coupling matrix defined in the
(s)lepton superpotential 
\begin{equation}
W_E = \epsilon_{ab}  (Y_E)_{ij} H_1^a L_i^{bo} \bar{E}_j^o,
\end{equation}
where the charged lepton interaction eigenstates 
$\{ e_L^o, e_R^o \}$ are related to the mass eigenstates
$\{ e_L, e_R, \}$ by 
\begin{equation}
e_L^o = V_e e_L \qquad \mbox{and} \qquad e_R^o = U_e e_R.
\end{equation}
The equivalent diagonalised charged lepton Yukawa matrix is 
\begin{equation}
(\hat{Y}_E)_{ii} = (U_e^\dag Y_E^T V_e)_{ii} = \sqrt{2}
  \frac{m_{ei}}{v_1}~~~. \label{eq:diagYL}
\end{equation}

Lepton mixing in the charged current interaction can then be characterised by
the PMNS matrix
\begin{equation}
U_{PMNS}=V_e^\dag V_\nu~, 
\end{equation}
which is proportional to the tree-level
$\bar{e}_{Li}{\nu_j}W^-$ and $\bar{e}_{Ri}{\nu_j}H^-$
couplings ($i,j=1,2,3$). This is also true for the supersymmetric
counterparts of these vertices, in the limit of unbroken supersymmetry.

Rotating the interaction eigenstates of the sleptons identically to
their leptonic counterparts, we obtain the super-PMNS basis for the charged
sleptons and the sneutrinos, described by the Lagrangian\footnote{We here 
neglect the possible term 
$\Phi_\nu^T {\hat {\mathcal M}}_{\tilde \nu}^2 \Phi_\nu$. }
\begin{equation}
{\mathcal L}_{\tilde l}^{mass} = - \Phi_e^\dag {\mathcal  M}_{\tilde
  e}^2 \Phi_e - 
\Phi_\nu^\dag {\mathcal M}_{\tilde \nu}^2 \Phi_\nu,
\end{equation}
where $\Phi_\nu=( {\tilde \nu}_e, {\tilde \nu}_\mu, {\tilde \nu}_\tau )^T$ and
$\Phi_e = ( {\tilde e}_L, {\tilde \mu}_L, {\tilde \tau}_L, {\tilde e}_R,
{\tilde \mu}_R, {\tilde \tau}_R )^T$. ${\mathcal M}_{\tilde e}^2$ is the $6 \times 6$ matrix
\begin{equation}
{\mathcal M}^2_{\tilde e} = \left( 
\begin{array}{cc} 
 {\hat{m}}^2_{\tilde L} + m_e^2 + {D_e}_{LL} & 
\frac{v_1}{\sqrt{2}} \hat{T}_E^\dag - \mu m_e \tan \beta \\[1.01ex]
\frac{v_1}{\sqrt{2}} \hat{T}_E - \mu^* m_e \tan \beta &
\hat{m}_{\tilde e}^2 + m_e^2 + {D_e}_{RR}
\end{array}
\right)~,
\end{equation}
and ${\mathcal M}_{\tilde \nu}^2$ is the $3\times 3$ matrix
\begin{equation}
{\mathcal M}_{\tilde \nu}^2 = U_{PMNS}^\dag \  \hat{m}_{\tilde L}^2 \ U_{PMNS}
+ {D_\nu}_{LL}, \label{numass}
\end{equation}
where  ${D_{e}}_{LL}$
and ${D_\nu}_{LL}$ are given in eq.~(\ref{dterm}).
In the equations above we introduced the $3 \times 3$ matrices
\begin{equation}
{\hat{m}_{\tilde L}}^2 \equiv V^\dagger_e \,m^2_{\tilde L}\, V_e\,,~~~
{\hat{m}_{\tilde e}}^2 \equiv U^\dagger_e \,{m^2_{\tilde e}}^T\, U_e\,,
\label{eq:mlhat}
\end{equation}
\begin{equation}
{\hat{T}_{E}} \equiv U^\dagger_e \,T_{E}^T\, V_e\,,~~~
\label{eq:tlhat}
\end{equation}
where the un-hatted mass matrices $m_{L,e}^2$ and 
the trilinear interaction matrix $T_E$ 
are given in the interaction basis. 
We diagonalise the charged slepton and sneutrino mass matrices via the
unitary 6$\times$6 and $3\times 3$
matrices $R_{e,\nu}$ respectively. Thus, $R_{e,\nu} {\mathcal M}^2_{\tilde e,
  \tilde \nu} R_{e,\nu}^\dagger$ are diagonal with increasing entries toward
the bottom right of each matrix. 

\subsubsection{Explicit proposal for SLHA2 \label{sec:flvproposal}}

As in the SLHA1~\cite{Skands:2003cj}, for all running parameters in
the output of the spectrum file, we propose to use definitions in the
modified dimensional reduction ($\overline{\mrm{DR}}$) 
scheme. 
The basis is the super-CKM/PMNS basis as defined above, that is the one 
in which the Yukawa couplings of the SM fermions, 
given in the $\overline{\mrm{DR}}$
scheme, are diagonal. Note that the masses and vacuum expectation
values (VEVs) in 
eqs.~(\ref{diagyukawa}), (\ref{eq:diagmL}), and (\ref{eq:diagYL}) 
must thus also be the running ones in the 
$\overline{\mrm{DR}}$ scheme. 

The input for an explicit implementation in a spectrum calculator 
consists of the following information:
\begin{itemize}
\item By default, all input SUSY parameters are given at the scale
  $M_\mrm{input}$ as defined in the SLHA1 block \ttt{EXTPAR} (see
  above). In principle, advanced codes may also allow for 
separate input scales for the sfermion mass matrices and trilinear
 couplings, via the block \ttt{QEXTPAR} defined above, 
 but we emphasise that this should be regarded as non-standard.
\item For the SM input parameters, we take
the Particle Data Group (PDG) definition: 
lepton masses are all on-shell. 
The light quark masses $m_{u,d,s}$ are given at 2 GeV, and the heavy
quark masses are given as 
$m_c(m_c)^{\overline{\mrm{MS}}}$, $m_b(m_b)^{\overline{\mrm{MS}}}$ and
$m_{t}^{\mrm{on-shell}}$. 
The latter two quantities are already in the SLHA1. The others are
added to \ttt{SMINPUTS} in the following manner (repeating the SLHA1
parameters for convenience):\\[2mm]
\numentry{1}{$\alpha_\mrm{em}^{-1}(m_{\Z})^{\MSbar}$. Inverse 
 electromagnetic coupling at the $\Z$ pole in the $\MSbar$ scheme  (with 5
 active flavours).} 
\numentry{2}{$G_F$. Fermi constant (in units of $\GeV^{-2}$).}
\numentry{3}{$\alpha_s(m_{\Z})^{\MSbar}$. Strong coupling at
  the $\Z$ pole in the $\MSbar$ scheme (with 5 active flavours).}
\numentry{4}{$m_\Z$, pole mass.}
\numentry{5}{$m_b(m_b)^{\MSbar}$. $b$ quark running mass in the $\MSbar$
  scheme.}
\numentry{6}{$m_t$, pole mass.}
\numentry{7}{$m_\tau$, pole mass.}
\numentry{8}{$m_{\nu_3}$, pole mass.}
\numentry{11}{$m_\e$, pole mass. }
\numentry{12}{$m_{\nu_1}$, pole mass.} 
\numentry{13}{$m_\mu$, pole mass. }
\numentry{14}{$m_{\nu_2}$, pole mass.} 
\numentry{21}{$m_d(2\ \GeV)^{\MSbar}$. $d$ quark running mass in the $\MSbar$
  scheme.}
\numentry{22}{$m_u(2\ \GeV)^{\MSbar}$. $u$ quark running mass in the $\MSbar$
  scheme.}
\numentry{23}{$m_s(2\ \GeV)^{\MSbar}$. $s$ quark running mass in the $\MSbar$
  scheme.}
\numentry{24}{$m_c(m_c)^{\MSbar}$. $c$ quark running mass in the $\MSbar$
  scheme.}
The FORTRAN 
format is the same as that of \ttt{SMINPUTS} in SLHA1 \cite{Skands:2003cj}.
\item $V_{\mrm{CKM}}$: the input CKM matrix in the Wolfenstein
parameterisation\footnote{For the Wolfenstein parameters we use the
PDG definition, eq.(11.4) of \cite{Yao:2006px}, which is exact to all orders in
$\lambda$.}, in the block \ttt{VCKMIN}. Note that present CKM studies
do not precisely define a renormalisation scheme for this matrix since
the electroweak effects that renormalise it are highly suppressed and
generally neglected. We therefore assume that the CKM elements given
by PDG (or by \tsc{UTFit}~\cite{Ciuchini:2000de} and
\tsc{CKMFitter}\cite{Charles:2004jd}, the main collaborations that
extract the CKM parameters) refer to SM $\overline{\mrm{MS}}$
quantities defined at $Q=m_Z$, to avoid any possible
ambiguity. \ttt{VCKMIN} should have the following entries \\[2mm]
 \numentry{1}{$\lambda$}
 \numentry{2}{$A$}
 \numentry{3}{$\bar{\rho}$}
 \numentry{4}{$\bar{\eta}$}
The FORTRAN format is the same as that of \ttt{SMINPUTS} above. 
\item $U_{\mrm{PMNS}}$: the input PMNS matrix, in the block
  \ttt{UPMNSIN}. It should have the PDG parameterisation in terms of
  rotation angles~\cite{Yao:2006px}  (all in radians):\\[2mm]
 \numentry{1}{$\bar \theta_{12}$ (the solar angle)}
 \numentry{2}{$\bar \theta_{23}$ (the atmospheric mixing angle)}
 \numentry{3}{$\bar \theta_{13}$ (currently only has an upper bound)}
 \numentry{4}{$\bar \delta_{13}$ (the Dirac CP-violating phase)}
 \numentry{5}{$\alpha_1$ (the first Majorana CP-violating phase)}
\numentry{6}{$\alpha_2$ (the second CP-violating Majorana phase)}
The FORTRAN format is the same as that of \ttt{SMINPUTS} above. Majorana
  phases have no effect on neutrino oscillations. However, they have physical
  consequences in the case of, for example, $\beta \beta 0 \nu$ decay of
  nuclei.
\item $(\hat{m}^2_{\tilde Q})_{ij}^{\overline{\mrm{DR}}}$,  
$(\hat{m}^2_{\tilde u})_{ij}^{\overline{\mrm{DR}}}$,
$(\hat{m}^2_{\tilde d})_{ij}^{\overline{\mrm{DR}}}$, 
$(\hat{m}^2_{\tilde L})_{ij}^{\overline{\mrm{DR}}}$,   
$(\hat{m}^2_{\tilde e})_{ij}^{\overline{\mrm{DR}}}$: the squark and
  slepton soft SUSY-breaking masses at the input scale 
in the super-CKM/PMNS basis, as defined above.
They will be given in the 
new blocks \ttt{MSQ2IN}, \ttt{MSU2IN},
\ttt{MSD2IN}, \ttt{MSL2IN}, \ttt{MSE2IN}, 
with the FORTRAN format 
\begin{verbatim}
(1x,I2,1x,I2,3x,1P,E16.8,0P,3x,'#',1x,A).
\end{verbatim}
where the first two integers in the format correspond to $i$ and 
$j$ and the double precision number to the soft mass squared. 
Only the ``upper
triangle'' of these matrices should be given. 
If diagonal entries are present, these supersede the 
parameters in the SLHA1 block \ttt{EXTPAR}. 
\item $(\hat{T}_U)_{ij}^{\overline{\mrm{DR}}}$, 
 $(\hat{T}_D)_{ij}^{\overline{\mrm{DR}}}$, and 
$(\hat{T}_E)_{ij}^{\overline{\mrm{DR}}}$: 
the squark and slepton  
soft SUSY-breaking trilinear couplings at the input scale 
in the super-CKM/PMNS basis. They will be given in the 
new blocks \ttt{TUIN}, \ttt{TDIN},
\ttt{TEIN}, in the same format as the soft mass
 matrices above. If diagonal entries are present these supersede
 the $A$ parameters specified in the SLHA1 block \ttt{EXTPAR} 
\cite{Skands:2003cj}.
\end{itemize}
For the output, the pole masses are given in block \ttt{MASS} as in
SLHA1 (note, however, that some PDG numbers have different
assignments in SLHA2, see below) 
and the $\DRbar$ and mixing parameters as follows:
\begin{itemize}
\item $(\hat{m}^2_{\tilde Q})_{ij}^{\overline{\mrm{DR}}}$,  
$(\hat{m}^2_{\tilde u})_{ij}^{\overline{\mrm{DR}}}$,
$(\hat{m}^2_{\tilde d})_{ij}^{\overline{\mrm{DR}}}$,
  $(\hat{m}^2_{\tilde L})_{ij}^{\overline{\mrm{DR}}}$,   
$(\hat{m}^2_{\tilde e})_{ij}^{\overline{\mrm{DR}}}$: the squark and
  slepton soft SUSY-breaking masses at scale $Q$ in the super-CKM/PMNS basis. 
Will be given in the 
new blocks \ttt{MSQ2 Q=...}, \ttt{MSU2 Q=...}, \ttt{MSD2 Q=...},
\ttt{MSL2 Q=...}, \ttt{MSE2 Q=...}, with 
formats as the corresponding input blocks \ttt{MSX2IN}
above. 
\item $(\hat{T}_U)_{ij}^{\overline{\mrm{DR}}}$, 
 $(\hat{T}_D)_{ij}^{\overline{\mrm{DR}}}$, and 
$(\hat{T}_E)_{ij}^{\overline{\mrm{DR}}}$: 
The squark and slepton 
soft SUSY-breaking trilinear couplings in the super-CKM/PMNS
 basis. Given in the new blocks \ttt{TU Q=...},  \ttt{TD Q=...},  
\ttt{TE Q=...}, which supersede the SLHA1 blocks \ttt{AD}, \ttt{AU},
 and \ttt{AE}, see \cite{Skands:2003cj}.
\item $(\hat{Y}_U)_{ii}^{\overline{\mrm{DR}}}$,
$(\hat{Y}_D)_{ii}^{\overline{\mrm{DR}}}$,
$(\hat{Y}_E)_{ii}^{\overline{\mrm{DR}}}$: the diagonal 
$\overline{\mrm{DR}}$ Yukawas
in the super-CKM/PMNS basis, with $\hat{Y}$ defined by
eqs.~(\ref{diagyukawa}) and (\ref{eq:diagYL}),
at the scale $Q$. Given in the SLHA1 blocks \ttt{YU Q=...},  \ttt{YD Q=...},  \ttt{YE Q=...},  see \cite{Skands:2003cj}. 
Note that although the SLHA1 blocks provide for off-diagonal
elements, only the 
diagonal ones will be relevant here, due to the CKM/PMNS rotation.
\item The entries of the $\DRbar$ CKM matrix at the
scale $Q$. The real and imaginary parts are given in \ttt{VCKM Q=...} and
\ttt{IMVCKM Q=...}, respectively. The format of the individual entries
is the same as for mixing matrices in the SLHA1. Note that the
complete matrix should be output, i.e., all entries should be
included. 
\item  The entries of the $\DRbar$ PMNS matrix at the
scale $Q$. The real and imaginary parts are given in \ttt{UPMNS Q=...} and
\ttt{IMUPMNS Q=...}, respectively, with
entries defined as for the $V_{CKM}$ output blocks above.
\item 
The squark and slepton masses and mixing matrices should be defined as
in the existing SLHA1, e.g.\ extending the $\tilde t$, $\tilde b$ and
$\tilde \tau$ mixing matrices to the 6$\times$6 case. More specifically,
the new blocks $R_u=$\ttt{USQMIX} $R_d=$\ttt{DSQMIX}, $R_e=$\ttt{SELMIX} and
the $3\times 3$ 
matrix for $R_\nu=$\ttt{SNUMIX} specify the composition of the mass eigenstates
in terms of the super-CKM/PMNS basis states according to the following
definitions:
\begin{equation}
\left(\begin{array}{c}
1000001 \\
1000003 \\
1000005 \\
2000001 \\
2000003 \\
2000005 \\
\end{array}\right) =
\left(\begin{array}{c}
\ti{d}_1 \\
\ti{d}_2 \\
\ti{d}_3 \\
\ti{d}_4 \\
\ti{d}_5 \\
\ti{d}_6 \\
\end{array}\right)_{\!\mathrm{mass-ordered}}\hspace*{-1cm} = 
\mbox{\ttt{DSQMIX}}_{ij} \left(\begin{array}{c}
\ti{d}_L \\
\ti{s}_L \\
\ti{b}_L \\
\ti{d}_R \\
\ti{s}_R \\
\ti{b}_R \\
\end{array}\right)_{\!\mathrm{super-CKM}}~,
\end{equation}
\begin{equation}
\left(\begin{array}{c}
1000002 \\
1000004 \\
1000006 \\
2000002 \\
2000004 \\
2000006 \\
\end{array}\right) =
\left(\begin{array}{c}
\ti{u}_1 \\
\ti{u}_2 \\
\ti{u}_3 \\
\ti{u}_4 \\
\ti{u}_5 \\
\ti{u}_6 \\
\end{array}\right)_{\!\mathrm{mass-ordered}}\hspace*{-1cm} = 
\mbox{\ttt{USQMIX}}_{ij} \left(\begin{array}{c}
\ti{u}_L \\
\ti{c}_L \\
\ti{t}_L \\
\ti{u}_R \\
\ti{c}_R \\
\ti{t}_R \\
\end{array}\right)_{\!\mathrm{super-CKM}}~.
\end{equation}
\begin{equation}
\left(\begin{array}{c}
1000011 \\
1000013 \\
1000015 \\
2000011 \\
2000013 \\
2000015 \\
\end{array}\right) =
\left(\begin{array}{c}
\ti{e}_1 \\
\ti{e}_2 \\
\ti{e}_3 \\
\ti{e}_4 \\
\ti{e}_5 \\
\ti{e}_6 \\
\end{array}\right)_{\!\mathrm{mass-ordered}}\hspace*{-1cm} = 
\mbox{\ttt{SELMIX}}_{ij} \left(\begin{array}{c}
\ti{e}_L \\
\ti{\mu}_L \\
\ti{\tau}_L \\
\ti{e}_R \\
\ti{\mu}_R \\
\ti{\tau}_R \\
\end{array}\right)_{\!\mathrm{super-PMNS}}~,
\end{equation}
\begin{equation}
\left(\begin{array}{c}
1000012 \\
1000014 \\
1000016 \\
\end{array}\right) =
\left(\begin{array}{c}
\ti{\nu}_1 \\
\ti{\nu}_2 \\
\ti{\nu}_3 \\
\end{array}\right)_{\!\mathrm{mass-ordered}}\hspace*{-1cm} = 
\mbox{\ttt{SNUMIX}}_{ij} \left(\begin{array}{c}
\ti{\nu}_{e} \\
\ti{\nu}_{\mu} \\
\ti{\nu}_{\tau} \\
\end{array}\right)_{\!\mathrm{super-PMNS}}~.
\end{equation}
{\bf Note!} A potential for inconsistency arises if the masses and
mixings are not calculated in the same way, e.g.\ if radiatively
corrected masses are used with tree-level mixing matrices. In this
case, it is possible that the radiative corrections to the masses
shift the mass ordering relative to the tree-level. This is especially
relevant when near-degenerate masses occur in the spectrum and/or when
the radiative corrections are large. In these cases, 
explicit care must be taken especially by the program writing the
spectrum, but also by the one reading it, to properly 
arrange the rows in
the order of the mass spectrum actually used.  

\item {\bf Optionally}, 
  we allow for the possibility of the scalar and pseudoscalar
  components of the sneutrinos to be treated separately. In this case,
  we define separate PDG codes and mixing matrices for the scalar and 
  pseudoscalar sneutrinos, as follows:
\begin{equation}
\left(\begin{array}{c}
1000012 \\
1000014 \\
1000016 \\
\end{array}\right) =
\left(\begin{array}{c}
\ti{\nu}_{1S} \\
\ti{\nu}_{2S} \\
\ti{\nu}_{3S} \\
\end{array}\right)_{\!\mathrm{mass-ordered}}\hspace*{-1cm} = 
\mbox{\ttt{SNSMIX}}_{ij} \left(\begin{array}{c}
\sqrt{2}\Re{\ti{\nu}_{e}} \\
\sqrt{2}\Re{\ti{\nu}_{\mu}} \\
\sqrt{2}\Re{\ti{\nu}_{\tau}} \\
\end{array}\right)_{\!\mathrm{super-PMNS}}~~,
\end{equation}
\begin{equation}
\left(\begin{array}{c}
1000017 \\
1000018 \\
1000019 \\
\end{array}\right) =
\left(\begin{array}{c}
\ti{\nu}_{1A} \\
\ti{\nu}_{2A} \\
\ti{\nu}_{3A} \\
\end{array}\right)_{\!\mathrm{mass-ordered}}\hspace*{-1cm} = 
\mbox{\ttt{SNAMIX}}_{ij} \left(\begin{array}{c}
\sqrt{2}\Im{\ti{\nu}_{e}} \\
\sqrt{2}\Im{\ti{\nu}_{\mu}} \\
\sqrt{2}\Im{\ti{\nu}_{\tau}} \\
\end{array}\right)_{\!\mathrm{super-PMNS}}~~.
\end{equation}
If present, \ttt{SNSMIX} and \ttt{SNAMIX} supersede \ttt{SNUMIX}. 

\end{itemize}

\subsection{R-Parity Violation \label{sec:rpv}}

We write the R-parity violating superpotential in the interaction
basis as
\bea
W_{\rpv} &=& \epsilon_{ab} \left[
\frac{1}{2} \lambda_{ijk} L_i^a L_j^b \bar{E}_k +
\lambda'_{ijk} L_i^a Q_j^{bx} \bar{D}_{kx}
- \kappa_i L_i^a H_2^b \right] \nonumber \\
&& + \frac{1}{2} \lambda''_{ijk} \epsilon_{xyz} \bar{U}_{i}^x
\bar{D}_{j}^y \bar{D}_{k}^z 
, \label{eq:Wrpv}
\eea
where $x,y,z=1,\ldots,3$ are fundamental SU(3)$_C$ indices and
$\epsilon_{xyz}$ is the totally antisymmetric tensor in 3 dimensions with
$\epsilon_{123}=+1$. 
In eq.~(\ref{eq:Wrpv}), $\lambda_{ijk}, \lambda'_{ijk}$ and $\kappa_i$ break
lepton number, whereas $\lambda''_{ijk}$ violate baryon
number.  
To ensure proton stability, either lepton number
conservation or baryon number conservation is usually still assumed,
resulting in either 
$\lambda_{ijk}=\lambda'_{ijk}=\kappa_{i}=0$ or $\lambda''_{ijk}=0$ for all
$i,j,k=1,2,3$. 

The trilinear R-parity violating terms in the soft SUSY-breaking potential
are 
\bea
V_{3,\rpv} &=& \epsilon_{ab} \left[
\frac12(T)_{ijk} {\tilde L}_{iL}^a {\tilde L}_{jL}^b {\tilde e}^*_{kR} +
(T')_{ijk} {\tilde L}_{iL}^a {\tilde Q}_{jL}^b {\tilde d}^*_{kR} \right] \nonumber
\\ 
&&+\frac12 (T'')_{ijk}\epsilon_{xyz} {\tilde u}_{iR}^{x*} {\tilde d}_{jR}^{y*} {\tilde d}_{kR}^{z*}
+ \mathrm{h.c.} \label{eq:trilinear}~~~.
\eea
Note that we do not factor 
out the $\lambda$ couplings (e.g.\ as in 
${T_{ijk}}/{\lambda_{ijk}}\equiv A_{\lambda,ijk}$) in order to avoid
potential problems with $\lambda_{ijk}=0$ but $T_{ijk}\ne 0$.
This usage is 
consistent with the convention for the R-conserving sector elsewhere in
this  report.

The bilinear R-parity violating soft terms (all lepton number
violating) are 
\beq
V_{2,\rpv} = -\epsilon_{ab}D_i {\tilde L}_{iL}^a H_2^b  + {\tilde
L}_{iaL}^\dag m_{{\tilde L}_i H_1}^2 H_1^a + \mathrm{h.c.}~.
\label{eq:bilinear}
\eeq

When lepton number is not conserved the sneutrinos may 
acquire vacuum expectation values (VEVs)
$\langle {\tilde \nu}_{e,\mu,\tau} \rangle \equiv v_{e, \mu,\tau}/\sqrt{2}$.
The SLHA1 defined the VEV $v$, 
which at tree level is equal to $2 m_Z / \sqrt{g^2 +
  {g'}^2}\sim246$ GeV; this is now generalised to  
\beq
v=\sqrt{v_1^2+v_2^2+v_{\e}^2+v_{\mu}^2+v_{\tau}^2}~.
\eeq
The addition of sneutrino VEVs allows for
various different definitions of $\tan
\beta$, but we here choose to keep the SLHA1 definition $\tan
\beta=v_2/v_1$. 

For input/output, we 
use the super-CKM/PMNS basis throughout, as defined in
section~\ref{sec:flv} with the following considerations specific to
the R-parity violating case.

Firstly, the $d$-quark mass matrices
are given by
\begin{eqnarray}
\sqrt{2}(m_d)_{ij} = (Y_D)_{ij} v_1 + \lambda'_{kij} v_k  \,\,.
\label{eq:dquarkmassrp}
\end{eqnarray}
where $v_k$ are the sneutrino VEVs. 
Secondly, in the lepton number violating case,  
the PMNS matrix can only be defined consistently 
by taking into account the 1-loop contributions induced by the
lepton-number violating couplings (see, e.g., \cite{Hirsch:2000ef}). 
We here restrict our attention to scenarios in which 
there are no right-handed neutrinos and, thus, neutrino
masses are generated solely by the lepton number violating couplings. 
In this case, the PMNS matrix is not an independent input but an
output. 

For definiteness, and to keep the changes with respect to the R-parity
conserving case as limited as possible, we define the super-CKM basis as
the one where the Yukawa couplings $Y_D$ and $Y_U$ are 
diagonal.
The PMNS basis is defined as the basis where $Y_E$ is diagonal and the 
loop-induced neutrino mass matrix is diagonalised. 
In this way one obtains a uniquely defined set of parameters:
\begin{eqnarray}
{\hat \lambda}_{ijk} &\equiv& \lambda_{rst} V_{\nu,ri} V_{e,sj} 
 U^\dagger_{e,tk}~,
 \label{eq:rotlam} \\
{\hat \lambda'}_{ijk} &\equiv& \lambda'_{rst}  V_{\nu,ri} V_{d,sj} 
U^\dagger_{d,tk}~, \\
{\hat \kappa}_i &\equiv&   \kappa_r V_{e,ri}~, \label{eq:rotbi} \\
{\hat \lambda''}_{ijk} &\equiv& \lambda''_{rst}  U^\dagger_{u,ri}
 U^\dagger_{d,sj} U^\dagger_{d,tk}~,
\label{eq:rotlampp}
\end{eqnarray}
where  the fermion mixing matrices are defined in section~\ref{sec:flv}.
The Lagrangian for the quark-slepton interactions then takes the following
form:
\begin{eqnarray}
{\cal L} = - {\hat \lambda}'_{ijk}  \tilde \nu_i \bar{d}_{Rk} d_{Lj}
 + {\hat \lambda'}_{rsk} U_{PMNS,ri}^\dagger V_{CKM,sj}^\dagger 
 \tilde l_{L,i} \bar{d}_{Rk} u_{Lj} \, + \mathrm{h.c.}~.
\end{eqnarray}
Similarly one obtains the soft SUSY breaking couplings in this basis
by replacing 
the superpotential quantities in
eqs.~(\ref{eq:rotlam})--(\ref{eq:rotlampp}) by the 
corresponding soft SUSY breaking couplings. In addition we define:
\begin{eqnarray}
\hat m^2_{\tilde L_i H_1} \equiv V^\dagger_{e,ir}m^2_{\tilde L_r H_1}~.
\end{eqnarray}

\subsubsection{Input/Output Blocks}

As mentioned above, we use the super-CKM/PMNS basis throughout, for
both superpotential and soft SUSY-breaking terms. This applies to both
input and output\footnote{A code may need to convert internally the
parameters to the interaction basis. In this case it must supply -- or
take as additional inputs -- the individual rotation matrices of quark
and lepton superfields entering
eqs.~(\ref{eq:rotlam})--(\ref{eq:rotlampp}).}.  The naming convention
for input blocks is {\tt BLOCK RV\#IN}, where the '{\tt \#}' character
represents the name of the relevant output block given below (thus,
for example, the ``LLE'' couplings in the super-PMNS basis,
$\hat\lambda_{ijk}$, would be given in {\tt BLOCK RVLAMLLEIN}).

Default inputs for all R-parity violating couplings are zero.
The inputs are given at scale $M_{\mrm{input}}$, as described 
in SLHA1 (again, if no $M_{\mrm{input}}$ is given, the GUT scale is
assumed), and follow the output format given below (with the omission of {\tt
  Q= ...}). In addition, the known fermion masses should be given in
\ttt{SMINPUTS} as defined in section \ref{sec:flvproposal}. 

The dimensionless super-CKM/PMNS 
couplings $\hat\lambda_{ijk}$, $\hat\lambda_{ijk}'$, and 
$\hat\lambda''_{ijk}$ are given in {\tt BLOCK RVLAMLLE, RVLAMLQD,
  RVLAMUDD Q= ...} respectively. The output standard should correspond to the
FORTRAN format 
\begin{verbatim}
(1x,I2,1x,I2,1x,I2,3x,1P,E16.8,0P,3x,'#',1x,A) .
\end{verbatim}
where the first three integers in the format correspond to $i$, $j$,
and $k$ and the double precision number is the coupling. 

$\hat{T}_{ijk}$, $\hat{T}'_{ijk}$, and $\hat{T}''_{ijk}$ 
are given in {\tt BLOCK RVTLLE, RVTLQD, RVTUDD Q= ...} in 
the same format as for the $\hat{\lambda}$ couplings above. 

The bilinear superpotential and soft SUSY-breaking terms $\hat{\kappa}_i$, 
$\hat{D}_i$, and $\hat m_{{\tilde L}_iH_1}^2$ and the sneutrino VEVs 
are given in {\tt BLOCK RVKAPPA, RVD, RVM2LH1, RVSNVEV Q= ...} respectively, in
the format  
\begin{verbatim}
(1x,I2,3x,1P,E16.8,0P,3x,'#',1x,A) .
\end{verbatim}

The input and output blocks for R-parity violating couplings are summarised
in Tab.~\ref{tab:rpvSummary}.
\begin{table}
\begin{center}\begin{tabular}{|l|l|l|}
\hline 
Input block & Output block  & data \\ \hline
 {\tt RVLAMLLEIN}& {\tt RVLAMLLE}&  $i$ $j$ $k$ $\hat\lambda_{ijk}$ \\
 {\tt RVLAMLQDIN}& {\tt RVLAMLQD}&  $i$ $j$ $k$ $\hat\lambda'_{ijk}$ \\
 {\tt RVLAMUDDIN}& {\tt RVLAMUDD}&  $i$ $j$ $k$ $\hat\lambda''_{ijk}$ \\
 {\tt RVTLLEIN}& {\tt RVTLLE}&  $i$ $j$ $k$ $\hat{T}_{ijk}$ \\
 {\tt RVTLQDIN}& {\tt RVTLQD}&  $i$ $j$ $k$ $\hat{T}'_{ijk}$ \\
 {\tt RVTUDDIN}& {\tt RVTUDD}&  $i$ $j$ $k$ $\hat{T}''_{ijk}$ \\
\hline
\multicolumn{3}{|c|}{NB: One of the following \ttt{RV...IN} blocks must be left out:}\\
\multicolumn{3}{|c|}{(which one up to user and RGE code)}\\
 {\tt RVKAPPAIN}& {\tt RVKAPPA}&  $i$ $\hat\kappa_{i}$ \\
 {\tt RVDIN}& {\tt RVD}&  $i$ $\hat{D}_{i}$ \\
 {\tt RVSNVEVIN}& {\tt RVSNVEV}&  $i$ $v_{i}$ \\
 {\tt RVM2LH1IN}& {\tt RVM2LH1}&  $i$ $\hat{m}_{{\tilde L}_{i}H_1}^2$ \\
\hline \end{tabular}
\caption{Summary of R-parity violating SLHA2 data blocks. 
All parameters are given in the Super-CKM/PMNS basis. 
Only 3 out of the last 4 blocks are independent. 
Which block to leave out of the input is in principle up to the
user, with the caveat that a given spectrum calculator may
not accept all combinations.
See text for a precise definition of the format. \label{tab:rpvSummary}} 
\end{center}
\end{table}
As for the R-conserving MSSM, the bilinear terms (both SUSY-breaking
and SUSY-respect\-ing ones, including $\mu$) and the VEVs are not
independent parameters. They become related by the condition of
electroweak symmetry breaking. Thus, in the SLHA1, one had the
possibility \emph{either} to specify $m_{H_1}^2$ and $m_{H_2}^2$
\emph{or} $\mu$ and $m_A^2$. This carries over to the \rpv\ case,
where not all the parameters in the input blocks \ttt{RV...IN} in
Tab.~\ref{tab:rpvSummary} can be given simultaneously.  
Specifically, of the last 4
blocks only 3 are independent. One block is determined by minimising
the Higgs-sneutrino potential.  We do not here insist on a particular
choice for which of \ttt{RVKAPPAIN}, \ttt{RVDIN}, \ttt{RVSNVEVIN}, and
\ttt{RVM2LH1IN} to leave out, but leave it up to the spectrum
calculators to accept one or more combinations.

\subsubsection{Particle Mixing}
In general, the
neutrinos mix with the neutralinos. This requires a change in the definition of
the $4\times 4$ neutralino mixing matrix $N$ to a $7\times 7$ matrix. 
The Lagrangian contains the (symmetric) neutrino/neutralino mass matrix as 
\begin{equation}
\mathcal{L}^{\mathrm{mass}}_{{\tilde \chi}^0} =
-\frac12{\tilde\psi^0}{}^T{\mathcal M}_{\tilde\psi^0}\tilde\psi^0 +
\mathrm{h.c.}~, 
\end{equation}
in the basis of 2--component spinors $\tilde\psi^0 =$
$( \nu_e, \nu_\mu, \nu_\tau, -i\tilde b, -i\tilde w^3,  
\tilde h_1, \tilde h_2)^T$. We define the
unitary $7 \times 7$ 
neutrino/neutralino mixing matrix $N$ (block {\tt RVNMIX}), such that:
\begin{equation}
-\frac12{\tilde\psi^0}{}^T{\mathcal M}_{\tilde\psi^0}\tilde\psi^0
= -\frac12\underbrace{{\tilde\psi^0}{}^TN^T}_{{{\tilde \chi}^0}{}^T}
\underbrace{N^*{\mathcal
    M}_{\tilde\psi^0}N^\dagger}_{\mathrm{diag}(m_{{\tilde \chi}^0})}
\underbrace{N\tilde\psi^0}_{{\tilde \chi}^0}~,  \label{eq:neutmass}
\end{equation}
where the 7 (2--component) generalised neutrinos ${\tilde
  \chi}^0=(\nu_1,...,\nu_7)^T$ are
defined strictly mass-ordered, i.e., with the 1$^{st}$,2$^{nd}$,3$^{rd}$
lightest corresponding to the mass entries for the PDG codes 
\ttt{12}, \ttt{14}, and \ttt{16}, and the four heaviest to the
 PDG codes \ttt{1000022}, \ttt{1000023}, \ttt{1000025}, and \ttt{1000035} (see
  also appendix \ref{sec:PDG}).

{\bf Note!} although these codes are normally associated with names
that imply a specific flavour content, such as code \ttt{12} being
$\nu_\e$ and so forth, it would be exceedingly complicated to maintain
such a correspondence in the context of completely general mixing,
hence we do not make any such association here. The flavour content of
each state, i.e., of each PDG number, is in general \underline{only}
defined by its corresponding entries in the mixing matrix
\ttt{RVNMIX}. Note, however, that the flavour basis is ordered so as
to reproduce the usual associations in the trivial case (modulo the
unknown flavour composition of the neutrino mass eigenstates).

In the limit of CP conservation, the default convention is that $N$ be
a real matrix and one or more of the mass eigenstates may
have an apparent negative mass. The minus sign may be removed by phase
transformations on $\tilde \chi^0_i\equiv\nu_i$ as explained in
SLHA1~\cite{Skands:2003cj}.

Charginos and charged leptons may also mix in the case of $L$-violation. 
In a similar spirit to the neutralino mixing, we define\footnote{Note
  that the absence of a factor $1/2$ on the r.h.s.~of
  eq.~(\ref{eq:LCharMass}) 
  corrects and supersedes the published version of this paper.}
\begin{equation}
\mathcal{L}^{\mathrm{mass}}_{{\tilde \chi}^+} =
-{\tilde\psi^-}{}^T{\mathcal M}_{\tilde\psi^+}\tilde\psi^+ +
\mathrm{h.c.}~, \label{eq:LCharMass}
\end{equation}
in the basis of 2--component spinors $\tilde\psi^- =$
$({e_L},{\mu_L},{\tau_L},-i\tilde w^-\!,\tilde h_1^-)^T$, $\tilde\psi^+ =$
$(\bar{e}_R,\bar{\mu}_R,\bar{\tau}_R,-i\tilde w^+\!, \tilde h_2^+)^T$, 
where $\tilde w^\pm = (\tilde w^1 \mp \tilde w^2) / \sqrt{2}$. Note
that in the limit of no \rpv\ the lepton fields are mass eigenstates.

 We define the
unitary $5 \times 5$
charged fermion mixing matrices $U,V$, blocks {\tt RVUMIX, RVVMIX}, 
such that:
\begin{equation}
-{\tilde\psi^-}{}^T{\mathcal M}_{\tilde\psi^+}\tilde\psi^+
= -\underbrace{{\tilde\psi^-}{}^TU^T}_{{{\kappa}^-}{}^T}
\underbrace{U^*{\mathcal
    M}_{\tilde\psi^+}V^\dagger}_{\mathrm{diag}(m_{{\tilde \chi}^+})}
\underbrace{V\tilde\psi^+}_{{\kappa}^+}~.  \label{eq:chargmass}
\end{equation}
The generalised charged leptons $\tilde \chi^- \equiv(e_1,e_2,e_3,e_4,e_5)$ 
are four-component Dirac fermions, and the left-handed and right-handed parts 
of $e_i$ are the two-component fermions $\kappa^-_i$ and $\bar{\kappa}^+_i$, 
respectively. They are defined as strictly mass ordered, i.e., 
with the 3 lightest states corresponding to the PDG codes \ttt{11},
\ttt{13}, and \ttt{15}, and the two heaviest to the codes
\ttt{1000024}, \ttt{1000037}.
As for neutralino mixing, the flavour
content of each state is in no way implied by its PDG number, but is
\underline{only} defined by its entries in  \ttt{RVUMIX} and
\ttt{RVVMIX}. Note, however, that the flavour basis is ordered so
as to reproduce the usual associations in the trivial case.  
For historical reasons, codes \ttt{11}, \ttt{13}, and
\ttt{15} pertain to the negatively charged field while codes
\ttt{1000024} and \ttt{1000037} pertain to the opposite charge. The
components of $\ti{\chi}^-$ in ``PDG notation'' would thus be
\ttt{(11,13,15,-1000024,-1000037)}. 
In the limit of CP
conservation, $U$ and $V$ are chosen to be real by default. 

R-parity violation via lepton number violation implies that the
 sneutrinos can mix with the Higgs bosons. In the limit of 
 CP conservation the CP-even (-odd) Higgs bosons mix with real (imaginary)
 parts of the sneutrinos. We write the neutral scalars as $\phi^0
 \equiv \sqrt{2} \Re{(H_1^0, H_2^0, {\tilde \nu}_e, {\tilde \nu}_\mu, {\tilde
\nu}_\tau)^T}$, with the mass term
\beq
{\mathcal L} = - \frac12 {\phi^0}^T {\mathcal M}_{\phi^0}^2 \phi^0~,
\eeq
where ${\mathcal M}_{\phi^0}^2$ is a $5 \times 5$ symmetric mass matrix. 
We define the orthogonal $5 \times 5$ 
mixing matrix $\aleph$ (block {\tt RVHMIX}) by
\begin{equation}
-{\phi^0}{}^T{\mathcal M}_{\phi^0}^2
\phi^0
= -\underbrace{{\phi^0}{}^T {\mathbf \aleph}^T}_{{{
      \Phi}^0}{}^T} 
\underbrace{{\mathbf \aleph}{\mathcal
    M}_{\phi^0}^2\aleph^T}_{\mathrm{diag}(m_{{ \Phi}^0}^2)}
\underbrace{{\mathbf \aleph}\phi^0}_{{ \Phi}^0}~,  
\label{eq:sneutmass}
\end{equation}
where $\Phi^0 \equiv (h^0_1, h^0_2, h^0_3, h^0_4, h^0_5)$ are the
neutral scalar mass eigenstates in strictly increasing mass order
(that is, we use the 
label $h$ for any neutral scalar mass eigenstate, regardless of
whether it is more ``Higgs-like'' or ``sneutrino-like''). The states
are numbered sequentially by the PDG codes 
\ttt{(25,35,1000012,1000014,1000016)}, regardless of flavour
content. The same 
convention will be followed below for the neutral pseudoscalars and
the charged scalars.  

We write the neutral pseudo-scalars 
as $\bar\phi^0 \equiv \sqrt{2} \Im{(H_1^0, H_2^0,
  {\tilde \nu}_e, {\tilde 
  \nu}_\mu, {\tilde \nu}_\tau)^T}$,
with the mass term
\beq
{\mathcal L} = - \frac12 {{\bar \phi}^0}{}^T {\mathcal M}_{{\bar
  \phi}^0}^2 {\bar \phi}^0~,
\eeq
where ${\mathcal M}_{\bar \phi^0}^2$ is a $5 \times 5$ symmetric mass
matrix. We define 
the $4 \times 5$ mixing matrix $\bar \aleph$ (block {\tt RVAMIX}) by
\begin{equation}
-{\bar \phi^0}{}^T{\mathcal M}_{\bar \phi^0}^2
\bar \phi^0
= -\underbrace{{\bar \phi^0}{}^T {\bar \aleph}^T}_{{{
      \bar \Phi}^0}{}^T} 
\underbrace{{\bar \aleph}{\mathcal
    M}_{\bar \phi^0}^2{\bar \aleph}^T}_{\mathrm{diag}(m_{{\bar \Phi}^0}^2)}
\underbrace{{\bar \aleph}\bar \phi^0}_{{\bar \Phi}^0}~,  
\label{eq:sneutmass2}
\end{equation}
where $\bar\Phi^0 \equiv (A^0_1, A^0_2, A^0_3, A^0_4)$ 
are the pseudoscalar mass eigenstates, again in strictly increasing
mass order. The states
are numbered sequentially by the PDG codes
\ttt{(36,1000017, 1000018,1000019)}, regardless of 
flavour composition. The Goldstone boson $G^0$
(the ``5th component'') has been explicitly left out and the 
4 rows of $\bar\aleph$ form a set of orthonormal vectors. 

If the blocks {\tt RVHMIX, RVAMIX} are present, they {\em supersede}\/ the
  SLHA1 {\tt ALPHA} variable/block.

The charged sleptons and charged Higgs bosons also mix in the $8 \times 8$ mass
squared matrix ${\mathcal M}^2_{\phi^\pm}$ by a $7 \times 8$ matrix $C$
(block {\tt RVLMIX}):
\beq
{\mathcal L}=- 
\underbrace{({H_1^-}^*, {H^+_2},\tilde{e}_{L_i}^*,\tilde{e}_{R_j}^*) C^\dagger}_{{\Phi^+}} 
\underbrace{C {\mathcal M}^2_{\phi^\pm} C^\dagger}_{ \mathrm{diag}({\mathcal M}^2_{\Phi^\pm})} 
C \left (
\begin{array}{c} {H_1^-}\\[1mm] {H_2^+}^* \\  \tilde{e}_{L_k} \\
\tilde{e}_{R_l} \end{array} \right )\, \label{eq:higgsslep}~,
\eeq
where $i,j,k,l\in \{1,2,3\}$, $\alpha,\beta\in \{1,\ldots,6\}$ and
$\Phi^+=\Phi^-{}^\dagger\equiv(h^+_1,h^+_2,h^+_3,h^+_4,h^+_5,h^+_6,h^+_7)$;
these states are numbered sequentially by the PDG codes
\ttt{(37,1000011,1000013,1000015, 2000011,2000013,2000015)},
regardless of flavour composition.  The Goldstone boson $G^+$ (the
``8th component'') has been explicitly left out and the 7 rows of $C$
form a set of orthonormal vectors.
  
R-parity violation may also generate contributions to down-squark
mixing via additional left-right mixing terms, 
\begin{eqnarray}
\frac{1}{\sqrt{2}} v_1 {\hat T}^\dagger_{D,ij} - \mu m_{d,i}
 \tan\beta \delta_{ij}
+ \frac{v_k}{\sqrt{2}} {\hat T}^\dagger_{\lambda',kij} 
\end{eqnarray}
where $v_k$ are the sneutrino vevs.
  However, this only mixes the six down-type squarks amongst
themselves and so is identical to the effects of flavour mixing.  This
is covered in section~\ref{sec:flv} (along with other forms of flavour
mixing).

\subsection{CP Violation \label{sec:cpv}}

When adding CP violation to the MSSM model parameters and mixing
matrices (for a recent review see, 
e.g., the CPNSH report \cite{CPV-Accomando:2006ga}),  
the SLHA1 blocks are understood to contain the real parts of the relevant
parameters. The imaginary parts should be provided with exactly the same
format, in a separate block of the same name but prefaced by {\tt IM}. 
The defaults for all imaginary parameters will be zero. Thus, for example, 
{\tt BLOCK IMAU, IMAD, IMAE, Q= ...} would describe the imaginary parts of the
trilinear soft SUSY-breaking scalar couplings. For input, {\tt BLOCK IMEXTPAR}
may be used to provide the relevant imaginary parts of soft SUSY-breaking
inputs.   
In cases where the definitions of the current paper supersede
the SLHA1 input and output blocks, completely equivalent statements
apply. 

One special case is the $\mu$ parameter. When the real part of $\mu$
is given in \ttt{EXTPAR 23}, the imaginary part should be given in
\ttt{IMEXTPAR 23}, as above. However, when $|\mu|$ 
is determined by the conditions for 
electroweak symmetry breaking, only the phase $\varphi_\mu$ is taken 
as an input parameter. In this case, SLHA2 generalises the entry
\ttt{MINPAR 4} to contain the cosine of the phase (as opposed to just 
$\mathrm{sign}(\mu)$ in SLHA1), and we further introduce a new
block \ttt{IMMINPAR} whose entry \ttt{4} gives the sine of the
phase, that is:
\subsection*{\ttt{BLOCK MINPAR}}
\numentry{4}{CP conserved: $\mathrm{sign}(\mu)$.\\ CP violated:
  $\cos\varphi_\mu=\Re{\mu}/|\mu|$. 
}
\subsection*{\ttt{BLOCK IMMINPAR}}
\numentry{4}{CP conserved: n/a.\\ CP violated:
  $\sin\varphi_\mu=\Im{\mu}/|\mu|$. 
}
Note that $\cos\varphi_\mu$ coincides with  
  $\mathrm{sign}(\mu)$ in the CP-conserving case. 

When CP symmetry is broken, quantum corrections cause mixing between
the CP-even and CP-odd Higgs states. 
Writing the neutral scalar interaction eigenstates as 
$\phi^0 \equiv \sqrt{2} (\Re{H_1^0},$
$\Re{H_2^0},$ $\Im{H_1^0},$ $\Im{H_2^0})^T$ 
we define the $3 \times 4$  mixing matrix $S$ 
(blocks {\tt CVHMIX} and {\tt IMCVHMIX}) by 
\begin{equation}
-{\phi^0}{}^T{\mathcal M}_{\phi^0}^2
 \phi^0
= -\underbrace{{ \phi^0}{}^T {S}^T}_{{{
       \Phi}^0}{}^T} 
\underbrace{{S}^*{\mathcal
    M}_{\phi^0}^2{S}^\dagger}_{\mathrm{diag}(m_{{ \Phi}^0}^2)}
\underbrace{{S}\phi^0}_{{ \Phi}^0}~,  
\label{eq:cvhmass}
\end{equation}
where $\Phi^0 \equiv (h_1^0,h_2^0,h_3^0)^T$ are the mass eigenstates; these states are numbered sequentially by the PDG codes
\ttt{(25,35,36)}, regardless of flavour composition. 
That is, even though the PDG
reserves code \ttt{36} for the CP-odd state, we do not maintain such a
labelling here, nor one that reduces to it. This means one does have to
exercise some caution when taking the CP conserving limit.

The matrix $S$ thus gives the decomposition of the three physical mass
eigenstates in terms of the four interaction eigenstates, all in one
go, with the Goldstone boson $G^0$ explicitly projected out and the 3
rows of $S$ forming a set of orthonormal vectors. 

For comparison, in the literature, the projecting-out of the
Goldstone boson is often done as a separate step, by first performing
a rotation by the angle $\beta$. (This is, for instance, the
prescription followed by \tsc{CPsuperH}
\cite{Lee:2003nta}).  In such an approach, our matrix
$S$ would be decomposed as:
\beq  
  S \phi^0 = 
\left( \begin{array}{cccc}
& & & 0 \\
\multicolumn{3}{c}{{\cal O}_{3\times3}} & 0 \\ 
& & & 0
\end{array}
 \right)
\underbrace{\left(\begin{array}{cccc}
1 & 0 & 0 & 0 \\
0 & 1 & 0 & 0 \\
0 & 0 & -\sin\beta & \cos\beta\\
0 & 0 & \cos\beta & \sin\beta
\end{array}\right) \phi^0}_{(\sqrt{2} \Re{H_1^0},
\sqrt{2} \Re{H_2^0},A_{\rm tree}^0,G_{\rm tree}^0)^T}~,
\eeq
where ${\cal O}_{3\times3}$ gives the decomposition of the three
physical mass eigenstates in terms of the intermediate basis
$\tilde{\phi}^0= (\sqrt{2}\Re{H_1^0},\sqrt{2}\Re{H_2^0},A_{\rm tree}^0)^T$,
with $A_{\rm tree}^0$ denoting the tree-level MSSM non-Goldstone
pseudoscalar mass eigenstate.  Note that a simple rotation by $\beta$
suffices to translate between the two conventions, so whichever is the
more practical can easily be used.

A second alternative convention,
e.g.\ adopted by \tsc{FeynHiggs}
\cite{Heinemeyer:1998yj,Frank:2006yh}, is to also rotate the CP-even states
by the angle $\alpha$ as part of the first step. In this case, our
matrix $S$ would be decomposed as:
\beq  
  S \phi^0 = 
\left( \begin{array}{cccc}
& & & 0 \\
\multicolumn{3}{c}{{\cal R}_{3\times3}} & 0 \\ 
& & & 0
\end{array}
 \right)
\underbrace{\left(\begin{array}{cccc}
-\sin\alpha & \cos\alpha & 0 & 0 \\
\cos\alpha & \sin\alpha & 0 & 0 \\
0 & 0 & -\sin\beta & \cos\beta\\
0 & 0 & \cos\beta & \sin\beta
\end{array}\right) \phi^0}_{(h^0,
H^0,A^0,G^0)^T_{\rm tree}}~,
\eeq
with $\alpha$ defined as the mixing angle in the CP-even Higgs sector
at tree-level and ${\cal R}_{3\times3}$ giving the
decomposition of the three physical mass eigenstates in terms of the
intermediate basis 
$\tilde{\Phi}^0=(h^0,H^0,A^0)^T_{\rm tree}$, that is in terms of the 
the tree-level mass eigenstates. In order to translate between $S$ and
${\cal R}_{3\times 3}$, the tree-level angle $\alpha$ would thus also
be needed. This should be
given in the SLHA1 output {\tt BLOCK ALPHA}:
\subsection*{\ttt{BLOCK ALPHA}}
\entry{CP conserved: $\alpha$; precise definition up to spectrum
  calculator, see SLHA1.\\ CP violated:
  $\alpha_{\mathrm{tree}}$. Must be accompanied by the matrix $S$, as
  described above, in the blocks \ttt{CVHMIX} and \ttt{IMCVHMIX}.
}

For the neutralino and chargino mixing matrices, the default convention 
in SLHA1 (and hence for the CP conserving case) 
is that they be real matrices. One or more mass eigenvalues may then
have an apparent negative sign, which can be removed by a
phase transformation on $\tilde \chi_i$ as explained in
SLHA1~\cite{Skands:2003cj}. When going to CPV, the reason for
introducing the negative-mass convention in the first place, namely
maintaining the mixing matrices strictly real, disappears. 
We therefore here take
all masses real and positive, with $N$, $U$, and $V$ complex. This does
lead to a nominal dissimilarity with SLHA1 in the limit of vanishing CP
violation, but we note that the explicit CPV switch in \ttt{MODSEL}
can be used to decide unambiguously which convention to follow.

\section{The Next-to-Minimal Supersymmetric SM\label{sec:nmssm}}

The first question to be addressed in defining universal
conventions for the next-to-minimal supersymmetric standard
model is just what field content and which
couplings this name should apply to. The field content
is already fairly well agreed upon; we shall here define the
next-to-minimal case as
having exactly the field content of the MSSM with the addition of one 
gauge-singlet chiral superfield. As to couplings and parameterisations,
several definitions exist in the literature (for a recent review see, 
e.g., the CPNSH report \cite{NMSSM-Accomando:2006ga}).  
Rather than adopting a particular one, or treating each special
case separately, below we choose instead to work at the most general
level. Any particular special
case can then be obtained by setting different combinations of
couplings to zero. For the time being, however, we do specialise to the
SLHA1-like case without CP violation, R-parity violation, or flavour
violation. Below, we shall use the acronym NMSSM for this class of
models, but we emphasise that we
understand it to relate to field content only, and not
to the presence or absence of specific couplings. 

\subsection{Conventions}

We write the most general CP conserving NMSSM superpotential as (extending the
notation of SLHA1):
\beq\label{eq:nmssmsup}
W_{NMSSM} = W_{MSSM} - \epsilon_{ab}\lambda {S} {H}^a_1 {H}^b_2 + \frac{1}{3}
\kappa {S}^3 + \frac12\mu' S^2 +\xi_F S \ , \eeq 
where $W_{MSSM}$ is the MSSM superpotential, eq.~(\ref{eq:wmssm}). 
A non-zero $\lambda$ in combination with a VEV $\left< S
\right>$ of the singlet generates a contribution to the effective 
$\mu$ term $\mu_\mathrm{eff}= \lambda \left< S
\right> + \mu$, where the MSSM $\mu$ term is normally assumed to be
zero in NMSSM constructions, 
yielding $\mu_{\mathrm{eff}}=\lambda \left< S \right>$. 
The sign of the $\lambda$ term in eq.~(\ref{eq:nmssmsup}) coincides with the
one in~\cite{Ellwanger:2005dv,Ellwanger:2004xm} 
where the Higgs doublet superfields
appear in opposite order. 
The remaining terms represent a general cubic 
potential for the singlet; $\kappa$ is dimensionless, $\mu'$ has
dimension of mass\footnote{Note that the factors $1/2$ in front of the
  $\mu'$ and $m_S'$ terms in eqs.~(\ref{eq:nmssmsup}) and
  (\ref{eq:nmssmsoft}), respectively, 
correct and supersede the published version of this
  paper.}, 
and $\xi_F$ has dimension of mass squared. 
The soft SUSY-breaking terms relevant to the NMSSM are
\beq\label{eq:nmssmsoft}
V_\mathrm{soft} = V_{2,MSSM} + V_{3,MSSM} + m_\mathrm{S}^2 | S |^2 +
(-\epsilon_{ab}\lambda A_\lambda {S} {H}^a_1 {H}^b_2 + 
\frac{1}{3} \kappa A_\kappa {S}^3  
+ \frac12 m_{S}'^2 S^2 +\xi_S S
+ \mathrm{h.c.}) \ , \eeq
where $V_{i,MSSM}$ are the MSSM soft terms defined in
eqs.~(\ref{eq:slha1v3soft}) and (\ref{eq:slha1v2soft}), 
and we have introduced the notation 
$m_{S}'^2 \equiv B'\mu'$. 

At tree level, there are thus 15 parameters (in addition to $m_Z$
which fixes the sum of the squared Higgs VEVs) 
that are 
relevant for the Higgs sector of the R-parity and CP-conserving NMSSM: 
\beq 
\tan\!\beta,\ \mu,\ m_{H_1}^2,\ m_{H_2}^2,\ m_3^2,\
\lambda,\  \kappa,\ A_{\lambda},\  A_{\kappa},\ \mu',\ m_{S}'^2,\ \xi_F,\
\xi_S,\ \lambda \left< S \right>,\ m_S^2~.\label{eq:nmssmpar}
\eeq
The minimisation of the effective 
potential imposes 3 conditions on these
parameters, such that only 12 of them can be considered
independent. We leave it up to each spectrum
calculator to decide on which combinations to accept. 
For the purpose of this accord, we note only that to specify a general
model exactly 12 parameters from eq.~(\ref{eq:nmssmpar}) should be
provided in the input, including explicit zeroes for parameters
desired ``switched off''. However, since
$\mu=m_3^2=\mu'=m_{S}'^2=\xi_F=\xi_S=0$ in the majority of phenomenological
constructions, for convenience we also allow for a six-parameter
specification in terms of the reduced parameter list:
\beq
\tan\!\beta,\ m_{H_1}^2,\ m_{H_2}^2,\ 
\lambda,\  \kappa,\ A_{\lambda},\  A_{\kappa},\
 \lambda \left< S \right>,\ m_S^2~.\label{eq:nmssmpar-reduced}
\eeq

To summarise, in addition to $m_Z$, the input to the accord should contain 
either 12 parameters
from the list given in eq.~(\ref{eq:nmssmpar}), including zeroes for parameters
not present in the desired model, or it should contain 6 parameters from
the list in eq.~(\ref{eq:nmssmpar-reduced}), in which case the
remaining 6 ``non-standard'' parameters, $\mu$, $m_3^2$, $\mu'$, $m_{S}'^2$,
$\xi_F$, and $\xi_F$, will be assumed to be zero;
in both cases 
the 3 unspecified parameters (as, e.g.,  
$m_{H_1}^2$, $m_{H_2}^2$, and $m_S^2$) are assumed to be determined by the
minimisation of the effective potential.

\subsection{Input/Output Blocks}

Firstly, as described above in section~\ref{sec:modelselection}, 
{\tt BLOCK MODSEL} should contain the switch 3 with value 1, corresponding to
the choice of the NMSSM particle content. 

Secondly, for the parameters that are also present in the MSSM, we
re-use the corresponding SLHA1 entries. That is, $m_Z$ should be given
in \ttt{SMINPUTS} entry \ttt{4} and $m_{H_1}^2, m_{H_2}^2$ can be
given in the \ttt{EXTPAR} entries \ttt{21} and \ttt{22}. $\tan\beta$
should either be given in \ttt{MINPAR} entry 3 (default) or
\ttt{EXTPAR} entry 25 (user-defined input scale), as in SLHA1.  If
$\mu$ should be desired non-zero, it can be given in \ttt{EXTPAR}
entry \ttt{23}. The corresponding soft parameter $m_3^2$ can be given
in \ttt{EXTPAR} entry \ttt{24}, in the form
$m_3^2/(\cos\beta\sin\beta)$, see \cite{Skands:2003cj}. The notation
$m_A^2$ that was used for that parameter in the SLHA1 is no longer
relevant in the NMSSM context, but by keeping the definition in terms
of $m_3^2$ and $\cos\beta\sin\beta$ unchanged, we maintain an
economical and straightforward correspondence between the two cases.

Further, new entries in {\tt BLOCK EXTPAR} have been defined 
for the NMSSM specific input parameters, as follows. As in the SLHA1,
these parameters are all given at the common scale 
\mgut, which can either be left up to the spectrum calculator or given
explicitly using \ttt{EXTPAR 0} or \ttt{QEXTPAR} 
(see section \ref{sec:slha1}):  

\subsection*{\ttt{BLOCK EXTPAR}}
\arrdes{Input parameters specific to the NMSSM 
(in addition to the entries defined in section \ref{sec:slha1})}
\numentry{61}{$\lambda(\mgut)$. Superpotential trilinear Higgs $SH_2H_1$ coupling.}
\numentry{62}{$\kappa(\mgut)$. Superpotential cubic $S$ coupling.} 
\numentry{63}{$A_\lambda(\mgut)$. Soft trilinear Higgs $SH_2H_1$ coupling.}
\numentry{64}{$A_\kappa(\mgut)$. Soft cubic $S$ coupling.} 
\numentry{65}{$\lambda \left< S \right>(\mgut)$. Vacuum expectation value of
  the singlet (scaled by $\lambda$). } 
\numentry{66}{$\xi_F(\mgut)$. Superpotential linear $S$ coupling.} 
\numentry{67}{$\xi_S(\mgut)$. Soft linear $S$ coupling.}
\numentry{68}{$\mu'(\mgut)$. Superpotential quadratic $S$ coupling.}
\numentry{69}{$m_{S}'^2(\mgut)$.  Soft quadratic $S$ coupling (sometimes
  denoted $\mu'B'$).} 
\numentry{70}{$m_S^2(\mgut)$. Soft singlet mass squared.}

{\bf Important note:} only 12 of the parameters listed in
eq.~(\ref{eq:nmssmpar}) should be given as input at any one time
(including explicit zeroes for parameters desired ``switched off''),
the remaining ones being determined by the minimisation of the
effective potential. Which combinations to accept is left up to the
individual spectrum calculator programs. Alternatively, for minimal
models, 6 parameters of those listed in
eq.~(\ref{eq:nmssmpar-reduced}) should be given.

For non-zero values, signs can be either positive or negative.  As
noted above, the meaning of the already existing entries \ttt{EXTPAR
23} and \ttt{24} (the MSSM $\mu$ parameter and corresponding soft
term) are maintained, which allows, in principle, for non-zero values
for both $\mu$ and $\left< S \right>$. The reason for choosing
$\lambda\left<S\right>$ rather than $\left< S \right>$ as input
parameter 65 is that it allows more easily to recover the MSSM limit
$\lambda$, $\kappa \to 0$, $\left< S \right> \to \infty$ with $\lambda
\left< S \right>$ fixed.

In the spectrum output, running NMSSM parameters corresponding to
the \ttt{EXTPAR} entries above can be given in the block \ttt{NMSSMRUN
Q=...}:

\subsection*{\ttt{BLOCK NMSSMRUN Q=...}}
Output parameters specific to the NMSSM, given in the \DRbar\ scheme, at the
scale $Q$. As in the SLHA1, several of these blocks may be given
simultaneously in the output, each then corresponding to a specific
scale. See 
corresponding entries in \ttt{EXTPAR} above for definitions.\\[3mm]
\numentry{1}{$\lambda(Q)^\DRbar$.}
\numentry{2}{$\kappa(Q)^\DRbar$.}
\numentry{3}{$A_\lambda(Q)^\DRbar$.}
\numentry{4}{$A_\kappa(Q)^\DRbar$.}
\numentry{5}{$\lambda\left<S\right>(Q)^\DRbar$.}
\numentry{6}{$\xi_F(Q)^\DRbar$.}
\numentry{7}{$\xi_S(Q)^\DRbar$.}
\numentry{8}{$\mu'(Q)^\DRbar$.}
\numentry{9}{$m_{S}'^2(Q)^\DRbar$.}
\numentry{10}{$m_S^2(Q)^\DRbar$.}

\subsection{Particle Mixing}

In the CP-conserving NMSSM, the CP-even interaction eigenstates are 
$\phi^0 \equiv\sqrt{2} \Re{(H_{1}^0, H_{2}^0, S)^T}$. 
We define the orthogonal $3 \times 3$ mixing matrix $S$ (block {\tt
  NMHMIX}) by 
\begin{equation}
-{\phi^0}{}^T{\mathcal M}_{\phi^0}^2 \phi^0
= -\underbrace{{\phi^0}{}^T {S}^T}_{{{
      \Phi}^0}{}^T} 
\underbrace{{S}{\mathcal
    M}_{\phi^0}^2{S}^T}_{\mathrm{diag}(m_{{\Phi}^0}^2)}
\underbrace{{S} \phi^0}_{{\Phi}^0}~,  
\end{equation}
where $\Phi^0 \equiv (h^0_1, h^0_2, h^0_3)$ are the mass eigenstates
ordered in mass. These states are
numbered sequentially by the PDG 
codes \ttt{(25,35,45)}. 
The format of {\tt BLOCK NMHMIX} is the same as for
the mixing matrices in SLHA1.

In the MSSM limit ($\lambda$, $\kappa \to 0$, and parameters such that
$h^0_3 \sim \Re{S}$) the elements of the first $2 \times 2$ sub-matrix of
$S_{ij}$ are related to the MSSM angle $\alpha$ as
\bea
S_{11} \sim &-\sin\alpha\ , \qquad &S_{21} \sim \cos\alpha\ ,\nonumber \\
S_{12} \sim &\cos\alpha\ , \qquad &S_{22} \sim \sin\alpha\ .\nonumber
\eea

In the CP-odd sector the interaction eigenstates are 
$\bar\phi^0 \equiv\sqrt{2} \Im{(H_{1}^0, H_{2}^0, S)^T}$. 
We define the $2 \times 3$ mixing matrix $P$ (block {\tt NMAMIX}) by
\begin{equation}
-{\bar \phi^0}{}^T{\mathcal M}_{\bar \phi^0}^2
\bar \phi^0
= -\underbrace{{\bar \phi^0}{}^T {P}^T}_{{{
      \bar \Phi}^0}{}^T} 
\underbrace{{P}{\mathcal
    M}_{\bar \phi^0}^2{P}^T}_{\mathrm{diag}(m_{{\bar \Phi}^0}^2)}
\underbrace{{P}\bar \phi^0}_{{\bar \Phi}^0}~,  
\end{equation}
where $\bar\Phi^0 \equiv (A^0_1, A^0_2)$ are the mass eigenstates
ordered in mass. These states are numbered sequentially by the PDG
codes \ttt{(36,46)}. The Goldstone boson $G^0$ (the ``3rd component'') has
been explicitly left out and the 2 rows of $P$ form a set of
orthonormal vectors.  An updated version \ttt{NMSSMTools}
\cite{Ellwanger:2005dv} will follow these conventions.

If {\tt NMHMIX, NMAMIX} blocks are present, they {\em supersede} the
SLHA1 {\tt ALPHA} variable/block.

The neutralino sector of the NMSSM requires a change in the definition
of the $4 \times 4$ neutralino mixing matrix $N$ to a $5 \times 5$ matrix.  The
Lagrangian contains the (symmetric) neutralino mass matrix as 
\begin{equation}
\mathcal{L}^{\mathrm{mass}}_{{\tilde \chi}^0} =
-\frac12{\tilde\psi^0}{}^T{\mathcal M}_{\tilde\psi^0}\tilde\psi^0 +
\mathrm{h.c.}~, 
\end{equation}
in the basis of 2--component spinors $\tilde\psi^0 =$ $(-i\tilde b,$
$-i\tilde w^3,$  $\tilde h_1,$ $\tilde h_2,$ $\tilde s)^T$. 
We define the unitary $5 \times 5$ neutralino mixing matrix $N$ (block {\tt
NMNMIX}), such that:
\begin{equation}\label{eq:nmssmneutmass}
-\frac12{\tilde\psi^0}{}^T{\mathcal M}_{\tilde\psi^0}\tilde\psi^0
= -\frac12\underbrace{{\tilde\psi^0}{}^TN^T}_{{{\tilde \chi}^0}{}^T}
\underbrace{N^*{\mathcal
    M}_{\tilde\psi^0}N^\dagger}_{\mathrm{diag}(m_{{\tilde \chi}^0})}
\underbrace{N\tilde\psi^0}_{{\tilde \chi}^0}~,  
\end{equation}
where the 5 (2--component) neutralinos ${\tilde \chi}_i$ are defined
such that the absolute value of their masses increase with $i$. As in
SLHA1, our convention is that $N$ be a real matrix. One or more mass
eigenvalues may then have an apparent negative sign, which can be
removed by a phase transformation on $\tilde \chi_i$. The states are
numbered sequentially by the PDG codes
\ttt{(1000022,1000023,1000025,1000035,1000045)}.

\section{Conclusion and Outlook}

At the time of writing of the SLHA1, a large number of computer codes
already existed which used MSSM spectrum and coupling information in
one form or another.  This had several advantages: there was a high
motivation from program authors to produce and implement the accord
accurately and quickly, and perhaps more importantly, the SLHA1 was
tested ``in anger'' in diverse situations as it was being written.

We find ourselves in a slightly different situation in terms of the
SLHA2.  There are currently few programs that utilise information in
any of the NMSSM or CP-violating, R-parity violating, or non-trivial
flavour violating MSSM scenarios.  Thus we do not have the benefit of
comprehensive simultaneous testing of the proposed accord and the
strong motivation that was present for implementation and writing of
the original one. What we do have are the lessons learned in
connection with the SLHA1 itself, and also several almost-finished
codes which are now awaiting the finalisation of SLHA2 in order to
publish their first official releases. Concrete tests involving
several of these were thus possible in connection with this writeup.

We have adhered to the principle of backward compatibility wherever
feasible.  We therefore expect that the conventions and agreements
reached within this paper constitute a practical solution that will
prove useful for SUSY particle phenomenology in the future.

\section*{Acknowledgements}
The majority of the agreements and conventions contained herein
resulted from the workshops ``Physics at TeV Colliders'', Les Houches,
France, 2005 \cite{Allanach:2006fy} and 2007, and ``Flavour in the Era
of the LHC'', CERN, 2005--2006. We also thank W.~Hollik for useful
discussions at ``Tools for SUSY and the New Physics'', 
Annecy-le-Vieux, France, 2006. 

BCA and WP would like to thank
enTapP 2005, Valencia, Spain, 2005 for hospitality offered during
working discussions of this project.  SM thanks The Royal
Society (London, UK) for partial financial support in the form of a
`Conference Grant' to attend the workshop ``Physics at TeV
Colliders'', Les Houches, France, 2007.

This work has been partially
supported by STFC and by Fermi Research Alliance, LLC, under Contract
No.\ DE-AC02-07CH11359 with the United States Department of
Energy. SP is supported by a Spanish MCyT Ramon y Cajal
contract. WP is supported by the German Ministry of Education and
Research (BMBF) under contract 05HT6WWA. 
The work of TG is supported in part by the Grant-in-Aid for
Science Research, Ministry of Education, Culture, Sports, Science and
Technology, Japan, No.\ 16081211. PG is supported by MIUR under
contract 2004021808-009. The work of MS is supported in part by the
Swiss Bundesamt f\"ur Bildung und Wissenschaft. The work of AP was
supported in part by GDRI-ACPP of CNRS and by 
grant RFBR-08-02-00856-a of the Russian Foundation for Basic Research.
JG, PG, T.~Hahn, SH, SP, and
WP are supported in part by the European Community's Marie-Curie
Research Training Network under contract MRTN-CT-2006-035505 `Tools
and Precision Calculations for Physics Discoveries at Colliders' and
PS is supported by contract MRTN-CT-2006-035606 `MCnet'. 

\clearpage\appendix
\section{PDG Codes and Extensions \label{sec:PDG}} 
The existing PDG nomenclature for (s)particle names is based
on the limit of the MSSM in which CP, R-parity, and flavour are
conserved. Several of the mass eigenstates 
are therefore labeled to indicate 
definite R, CP, and/or flavour quantum numbers. When the corresponding
symmetries are broken, such a labeling becomes
misleading. Throughout this paper we have adopted the convention of assigning
a common label to all states which carry identical conserved quantum
numbers in the given model. We then re-use the existing PDG codes for
those states, arranged in strictly increasing mass order. 

This implies that, while the PDG numbers remain unaltered, their
labels change, depending on which scenario is considered. The PDG
codes and labels are discussed in detail in the individual sections on
flavour violation, R-parity violation, CP violation, and the NMSSM. In
the tables below, we summarise the PDG numbers and suggested
labels relevant to each distinct scenario, for squarks 
(Tab.~\ref{tab:pdg1}), charged colour-singlet fermions (Tab.~\ref{tab:pdg2}),
neutral colour-singlet fermions (Tab.~\ref{tab:pdg22}), 
charged colour-singlet scalars (Tab.~\ref{tab:pdg3}), and neutral
colour-singlet scalars (Tab.~\ref{tab:pdg4}), respectively. Note that these 
extensions are not officially endorsed by the PDG at this
time. Codes for other particles can be found in \cite[chp.~33]{Yao:2006px}.

\begin{table}[hpt]
\arrdes{Scalar Quarks}
\center
\begin{tabular}{|c|c|c|c|c|c|c||c|}
\hline
FLV &\rbox{No} &\gbox{Yes}&\rbox{No} &\rbox{No}
&\gbox{Yes}&\gbox{Yes}&\\[-1mm]
RPV &\rbox{No} &\rbox{No} &\gbox{Yes}&\rbox{No} &\gbox{Yes}&\rbox{No}&\\[-1mm]
CPV &\rbox{No} &\rbox{No} &\rbox{No} &\gbox{Yes}&\rbox{No} &\gbox{Yes}&\\[-24mm]
& & & & & & & \rotatebox{270}{\hspace*{-2mm}\bbox{$\!$NMSSM}\ }\\[13mm]
\hline
& & & & & & & \\[-4mm]
1000001 & $\ti{d}_L$ & $\ti{d}_1$ & $\ti{d}_1$ & $\ti{d}_L$ &
$\ti{d}_1$ & $\ti{d}_1$ & $\ti{d}_L$\\
1000002 & $\ti{u}_L$ & $\ti{u}_1$ & $\ti{u}_1$ & $\ti{u}_L$ &
$\ti{u}_1$ & $\ti{u}_1$ & $\ti{u}_L$\\
1000003 & $\ti{s}_L$ & $\ti{d}_2$ & $\ti{d}_2$ & $\ti{s}_L$ &
$\ti{d}_2$ & $\ti{d}_2$ & $\ti{s}_L$\\
1000004 & $\ti{c}_L$ & $\ti{u}_2$ & $\ti{u}_2$ & $\ti{c}_L$ &
$\ti{u}_2$ & $\ti{u}_2$ & $\ti{c}_L$\\
1000005 & $\ti{b}_1$ & $\ti{d}_3$ & $\ti{d}_3$ & $\ti{b}_1$ &
$\ti{d}_3$ & $\ti{d}_3$ & $\ti{b}_1$\\
1000006 & $\ti{t}_1$ & $\ti{u}_3$ & $\ti{u}_3$ & $\ti{t}_1$ &
$\ti{u}_3$ & $\ti{u}_3$ & $\ti{t}_1$\\
2000001 & $\ti{d}_R$ & $\ti{d}_4$ & $\ti{d}_4$ & $\ti{d}_R$ &
$\ti{d}_4$ & $\ti{d}_4$ & $\ti{d}_R$\\
2000002 & $\ti{u}_R$ & $\ti{u}_4$ & $\ti{u}_4$ & $\ti{u}_R$ &
$\ti{u}_4$ & $\ti{u}_4$ & $\ti{u}_R$\\
2000003 & $\ti{s}_R$ & $\ti{d}_5$ & $\ti{d}_5$ & $\ti{s}_R$ &
$\ti{d}_5$ & $\ti{d}_5$ & $\ti{s}_R$\\
2000004 & $\ti{c}_R$ & $\ti{u}_5$ & $\ti{u}_5$ & $\ti{c}_R$ &
$\ti{u}_5$ & $\ti{u}_5$ & $\ti{c}_R$\\
2000005 & $\ti{b}_2$ & $\ti{d}_6$ & $\ti{d}_6$ & $\ti{b}_2$ &
$\ti{d}_6$ & $\ti{d}_6$ & $\ti{b}_2$\\
2000006 & $\ti{t}_2$ & $\ti{u}_6$ & $\ti{u}_6$ & $\ti{t}_2$ &
$\ti{u}_6$ & $\ti{u}_6$ & $\ti{t}_2$\\
\hline
\end{tabular}
\caption{Particle codes and corresponding labels for squarks. 
  The labels in the first
  column correspond to the current PDG nomenclature.\label{tab:pdg1}}
\end{table}

\begin{table}[h]
\arrdes{Charged Leptons and Charginos}
\center
\begin{tabular}{|c|c|c|c|c|c|c||c|}
\hline
FLV &\rbox{No} &\gbox{Yes}&\rbox{No} &\rbox{No}
&\gbox{Yes}&\gbox{Yes}&\\[-1mm]
RPV &\rbox{No} &\rbox{No} &\gbox{Yes}&\rbox{No} &\gbox{Yes}&\rbox{No}&\\[-1mm]
CPV &\rbox{No} &\rbox{No} &\rbox{No} &\gbox{Yes}&\rbox{No} &\gbox{Yes}&\\[-24mm]
& & & & & & & \rotatebox{270}{\hspace*{-2mm}\bbox{$\!$NMSSM}\ }\\[13mm]
\hline
11 &$e^-$   &$e^-$   &$e^-_1$ & $e^-$  & $e^-_1$&$e^-$
& $e^-$\\
13 &$\mu^-$   &$\mu^-$   &$e^-_2$ & $\mu^-$  & $e^-_2$&$\mu^-$
& $\mu^-$\\
15 &$\tau^-$   &$\tau^-$   &$e^-_3$ & $\tau^-$  & $e^-_3$&$\tau^-$
& $\tau^-$\\
1000024 & $\charg_1$ & $\charg_1$ & $e^+_4$ & $\charg_1$ & $e^+_4$ & $\charg_1$ & $\charg_1$ \\   
1000037 & $\charg_2$ & $\charg_2$ & $e^+_5$ & $\charg_2$ & $e^+_5$ & $\charg_2$ & $\charg_2$ \\
\hline
\end{tabular}
\caption{Particle codes and corresponding labels for 
  charged colour-singlet fermions. The labels in the first column correspond
  to the current PDG nomenclature. Note that, for historical reasons, 
  codes \ttt{11}, \ttt{13}, and
  \ttt{15} pertain to negatively charged fields while codes
  \ttt{1000024} and \ttt{1000037} pertain to the opposite charge.
  \label{tab:pdg2}}
\end{table}

\begin{table}[h]
\arrdes{Neutrinos and Neutralinos}
\center
\begin{tabular}{|c|c|c|c|c|c|c||c|}
\hline
FLV &\rbox{No} &\gbox{Yes}&\rbox{No} &\rbox{No}
&\gbox{Yes}&\gbox{Yes}&\\[-1mm]
RPV &\rbox{No} &\rbox{No} &\gbox{Yes}&\rbox{No} &\gbox{Yes}&\rbox{No}&\\[-1mm]
CPV &\rbox{No} &\rbox{No} &\rbox{No} &\gbox{Yes}&\rbox{No} &\gbox{Yes}&\\[-24mm]
& & & & & & & \rotatebox{270}{\hspace*{-2mm}\bbox{$\!$NMSSM}\ }\\[13mm]
\hline
12 &$\nu_e$   &$\nu_1$   &$\nu_1$ & $\nu_e$  & $\nu_1$&$\nu_1$
& $\nu_e$\\
14 &$\nu_\mu$   &$\nu_2$   &$\nu_2$ & $\nu_\mu$  & $\nu_2$&$\nu_2$
& $\nu_\mu$\\
16&$\nu_\tau$   &$\nu_3$   &$\nu_3$ & $\nu_\tau$  & $\nu_3$&$\nu_3$
& $\nu_\tau$\\
1000022 & $\neut_1$ & $\neut_1$ & $\nu_4$ & $\neut_1$ & $\nu_4$ &
$\neut_1$ & $\neut_1$\\
1000023 & $\neut_2$ & $\neut_2$ & $\nu_5$ & $\neut_2$ & $\nu_5$ &
$\neut_2$ & $\neut_2$\\
1000025 & $\neut_3$ & $\neut_3$ & $\nu_6$ & $\neut_3$ & $\nu_6$ &
$\neut_3$ & $\neut_3$\\
1000035 & $\neut_4$ & $\neut_4$ & $\nu_7$ & $\neut_4$ & $\nu_7$ &
$\neut_4$ & $\neut_4$\\
1000045 & - & - & - & - & - & - & $\neut_5$\\
\hline
\end{tabular}
\caption{Particle codes and corresponding labels for 
   neutral colour-singlet fermions. The labels in the first column correspond
  to the current PDG nomenclature. 
  \label{tab:pdg22}}
\end{table}

\begin{table}[hpt]
\arrdes{Charged Higgs Boson and Charged Scalar Leptons}
\center
\begin{tabular}{|c|c|c|c|c|c|c||c|}
\hline
FLV &\rbox{No} &\gbox{Yes}&\rbox{No} &\rbox{No}
&\gbox{Yes}&\gbox{Yes}&\\[-1mm]
RPV &\rbox{No} &\rbox{No} &\gbox{Yes}&\rbox{No} &\gbox{Yes}&\rbox{No}&\\[-1mm]
CPV &\rbox{No} &\rbox{No} &\rbox{No} &\gbox{Yes}&\rbox{No} &\gbox{Yes}&\\[-24mm]
& & & & & & & \rotatebox{270}{\hspace*{-2mm}\bbox{$\!$NMSSM}\ }\\[13mm]
\hline
& & & & & & & \\[-4mm]
37 &$H^+$  & $H^+$ & $h^+_1$ & $H^+$ & $h^+_1$ & $H^+$ & $H^+$\\
1000011 & $\ti{e}^+_L$ & $\ti{e}^+_1$ & $h_2^+$ & $\ti{e}^+_L$ &
$h_2^+$ & $\ti{e}^+_1$ &$\ti{e}^+_L$ \\
1000013 & $\ti{\mu}^+_L$ & $\ti{e}^+_2$ & $h_3^+$ & $\ti{\mu}^+_L$ &
$h_3^+$ & $\ti{e}^+_2$ &$\ti{\mu}^+_L$ \\
1000015 & $\ti{\tau}^+_1$ & $\ti{e}^+_3$ & $h_4^+$ & $\ti{\tau}^+_1$ &
$h_4^+$ & $\ti{e}^+_3$ &$\ti{\tau}^+_1$ \\
2000011 & $\ti{e}^+_R$ & $\ti{e}^+_4$ & $h_5^+$ & $\ti{e}^+_R$ &
$h_5^+$ & $\ti{e}^+_4$ &$\ti{e}^+_R$ \\
2000013 & $\ti{\mu}^+_R$ & $\ti{e}^+_5$ & $h_6^+$ & $\ti{\mu}^+_R$ &
$h_6^+$ & $\ti{e}^+_5$ &$\ti{\mu}^+_R$ \\
2000015 & $\ti{\tau}^+_2$ & $\ti{e}^+_6$ & $h_7^+$ & $\ti{\tau}^+_2$ &
$h_7^+$ & $\ti{e}^+_6$ &$\ti{\tau}^+_2$ \\
\hline
\end{tabular}
\caption{Particle codes and corresponding labels for charged
  colour-singlet scalars. The labels in the first column correspond to
  the current PDG nomenclature. 
 \label{tab:pdg3}} 
\end{table}

\begin{table}[hpt]
\arrdes{Neutral Higgs Bosons and Scalar Neutrinos}
\center
\begin{tabular}{|c|c|c|c|c|c|c||c|}
\hline
FLV &\rbox{No} &\gbox{Yes}&\rbox{No} &\rbox{No}
&\gbox{Yes}&\gbox{Yes}&\\[-1mm]
RPV &\rbox{No} &\rbox{No} &\gbox{Yes}&\rbox{No} &\gbox{Yes}&\rbox{No}&\\[-1mm]
CPV &\rbox{No} &\rbox{No} &\rbox{No} &\gbox{Yes}&\rbox{No} &\gbox{Yes}&\\[-24mm]
& & & & & & & \rotatebox{270}{\hspace*{-2mm}\bbox{$\!$NMSSM}\ }\\[13mm]
\hline
& & & & & & & \\[-4mm]
25 &$h^0$  & $h^0$ & $h_1^0$ & $h^0_1$ & $h^0_1$ & $h^0_1$ & $h^0_1$\\
35 &$H^0$  & $H^0$ & $h_2^0$ & $h_2^0$ & $h_2^0$ & $h_2^0$ & $h_2^0$ \\
36 &$A^0$  & $A^0$ & $A_1^0$ & $h_3^0$ & $A_1^0$ & $h_3^0$ & $A_1^0$\\
45 & -      & -& -&- &- &- & $h_3^0$\\
46 & -      & -& -&- &- &- & $A_2^0$\\
1000012 & $\ti{\nu}_{e_L}$    &$\ti{\nu}_1$ ($\ti{\nu}_{1S}$)& $h_3^0$
&$\ti{\nu}_{e_L}$   & $h_3^0$ & $\ti{\nu}_1$&$\ti{\nu}_{e_L}$ \\
1000014 & $\ti{\nu}_{\mu_L}$ &$\ti{\nu}_2$ ($\ti{\nu}_{2S}$) & $h_4^0$
&$\ti{\nu}_{\mu_L}$ & $h_4^0$ &$\ti{\nu}_2$ &$\ti{\nu}_{\mu_L}$ \\
1000016 & $\ti{\nu}_{\tau_L}$&$\ti{\nu}_3$ ($\ti{\nu}_{3S}$) & $h_5^0$
&$\ti{\nu}_{\tau_L}$& $h_5^0$ & $\ti{\nu}_3$&$\ti{\nu}_{\tau_L}$ \\
1000017 & -             &($\ti{\nu}_{1A}$)&$A^0_2$&-&$A^0_2$&-&-\\ 
1000018 & -             &($\ti{\nu}_{2A}$)&$A^0_3$&-&$A^0_3$&-&-\\
1000019 & -             &($\ti{\nu}_{3A}$)&$A^0_4$&-&$A^0_4$&-&-\\
\hline
\end{tabular} 
\caption{Particle codes and corresponding labels for neutral
  colour-singlet scalars. The labels in the first column correspond to
  the current PDG nomenclature. The labels in parenthesis denote the optional 
separation of sneutrinos into separate scalar and
  pseudoscalar components.\label{tab:pdg4}} 
\end{table}

\clearpage
\bibliography{slha2_52}

\end{document}